\documentclass[submit]{bioRxiv}

\usepackage{lipsum}
\usepackage[authoryear]{natbib}
\setcitestyle{square}
\setcitestyle{numbers}
\setcitestyle{citesep={,}}
\begin{document}

 \leadauthor{Park}

\title{Sound reconstruction from human brain activity via a generative model with brain-like auditory features}
\shorttitle{Neural sound reconstruction}
\author[1,2,A,\Letter]{Jong-Yun Park \orcidlink{0009-0001-7013-2098}}
\author[2]{Mitsuaki Tsukamoto }
\author[1]{Misato Tanaka \orcidlink{0000-0002-5633-5630}}
\author[1,2,\Letter]{Yukiyasu Kamitani \orcidlink{0000-0002-9300-8268}}
\affil[1]{Department of Intelligence Science and Technology, Graduate School of Informatics, Kyoto University, Kyoto, Japan }
\affil[2]{Department of Neuroinformatics, ATR Computational Neuroscience Laboratories, Kyoto, Japan}
\affil[A]{Current address: Department of Psychiatry and behavioral sciences, Graduate School of Medical and Dental Sciences, Tokyo Medical and Dental University, Tokyo, Japan}
\date{}

\bigskip

\maketitle

\begin{abstract}
The successful reconstruction of perceptual experiences from human brain activity has provided insights into the neural representations of sensory experiences. However, reconstructing arbitrary sounds has been avoided due to the complexity of temporal sequences in sounds and the limited resolution of neuroimaging modalities. To overcome these challenges, leveraging the hierarchical nature of brain auditory processing could provide a path toward reconstructing arbitrary sounds. Previous studies have indicated a hierarchical homology between the human auditory system and deep neural network (DNN) models. Furthermore, advancements in audio-generative models enable to transform compressed representations back into high-resolution sounds. In this study, we introduce a novel sound reconstruction method that combines brain decoding of auditory features with an audio-generative model. Using fMRI responses to natural sounds, we found that the hierarchical sound features of a DNN model could be better decoded than spectrotemporal features. We then reconstructed the sound using an audio transformer that disentangled compressed temporal information in the decoded DNN features. Our method shows unconstrained sounds reconstruction capturing sound perceptual contents and quality and generalizability by reconstructing sound categories not included in the training dataset. Reconstructions from different auditory regions remain similar to actual sounds, highlighting the distributed nature of auditory representations. To see whether the reconstructions mirrored actual subjective perceptual experiences, we performed an experiment involving selective auditory attention to one of overlapping sounds. The results tended to resemble the attended sound than the unattended. These findings demonstrate that our proposed model provides a means to externalize experienced auditory contents from human brain activity. 

\end{abstract}


\begin{corrauthor}
park\_jy.psyc@tmd.ac.jp, kamitani@i.kyoto-u.ac.jp
\end{corrauthor}

\section*{Introduction}
The intersection of progress in neuroimaging and machine learning has given rise to new understandings of sensory neural representations by enabling the reconstruction of perceptual experiences from human brain activity. The primary efforts have been focused on the brain decoding of visual content, encompassing not only what people see but also extending to imagination and dreams 	\cite{Kamitani2005Decoding,Haxby2001Distributed,Harrison2009Decoding,Cichy2012Imagery,Horikawa2013Neural}. Recent research has leveraged deep neural network (DNN) techniques to reconstruct visual content perceived from functional magnetic resonance imaging (fMRI) responses \cite{Güçlütürk2017Reconstructing,Han2019Variational,Shen2019End-to-End,Shen2019Deep}. This use of image reconstruction from neural responses has allowed researchers not only to externalize visual imagery recalled from subjective memory \cite{Shen2019Deep} but also to capture aspects of the top-down process of visual perception, such as attention \cite{Horikawa2022Attention} and illusory experience \cite{Cheng2022Reconstruction}. The success achieved in visual reconstruction has underscored the potential of DNNs in interpreting and translating intricate neural activities.
 
In parallel, researchers have sought to apply similar principles to the auditory system to unravel the complex neural mechanisms underpinning our auditory experiences \cite{Santoro2017Reconstructing,Pasley2012Reconstructing,Défossez2022Decoding,Akbari2019Towards,Wang2018Reconstructing,Moses2019Real-time,Anumanchipalli2019Speech}. However, unlike visual reconstruction, sound decoding studies typically avoid reconstruction under unconstrained conditions for arbitrary sounds. This is primarily due to the broad diversity and complex temporal sequencing of sounds, coupled with the relatively low resolution of neuroimaging modalities. Traditionally, neuroimaging modalities such as electroencephalography (EEG) and magnetoencephalography (MEG) have been favored for auditory decoding due to their superior temporal resolution, as they capture real-time electrical activity from the scalp or sensors placed on the head. Nevertheless, their utilization has been confined to classifying predefined speech \cite{Moses2019Real-time,Chakrabarti2015Progress,Martin2018Decoding,Pei2011Decoding} and reconstructing constrained examples such as digits \cite{Akbari2019Towards} and several words \cite{Wang2018Reconstructing}. Additionally, due to its inherent temporal resolution constraints, fMRI has been limited to classification approaches \cite{Formisano2008"Who",Correia2015Decoding}.

Contrary to this common practice, recent studies suggest that reconstruction of unconstrained sound might be possible without requiring a precise alignment of temporal resolution between neural recordings and auditory stimuli.	One approach involves using the spatial patterns of fMRI to compensate for its limited temporal resolution, enabling the prediction of intricate temporal information. Santoro et al. \cite{Santoro2017Reconstructing} built a computational model to decode the physical features of natural sounds using high spatial resolution 7T fMRI responses. This model deployed many multivariate decoders to predict spectral-temporal modulation features from fMRI activation patterns. Impressively, these trained decoders could predict fine modulation changes from fMRI's coarse temporal sampling (2.6 s). To aid the interpretation of the decoded results, the study transformed the decoded features back into sounds. Despite the promising results, the reconstructed sounds lacked complex spectro-temporal patterns, resulting in temporally smoothed reconstructions, which posed recognition challenges for human listeners.
 
In alignment with these trends, researchers have also focused on the hierarchical auditory features that are processed in a manner analogous to the human brain. This is supported by evidence of a similar hierarchical processing structure in both the human auditory system and DNN models. Kell et al. \cite{Kell2018Task-Optimized} carried out a comprehensive brain encoding analysis, where they predicted human auditory responses from the DNN model responses, showcasing the hierarchical homology between the DNN model and fMRI data. They crafted a DNN architecture for sound recognition, designed to mirror the hierarchical processing integral to the human auditory system. This DNN's structure is characterized by the segregation of common layers, tasked with low-level processing akin to early auditory stages, from branching layers that handle task-specific processing, such as speech recognition or music genre classification. Using this trained DNN model and fMRI data for natural sounds, their encoding analysis indicated a hierarchical correspondence between the brain and DNN models, with early auditory cortex brain responses better predicted from DNN features from the common layer, while brain responses of nonprimary regions were better predicted from DNN features from the branched layer. However, a subsequent study observed that not all DNNs exhibit homology with the human brain, as their encoding performance can vary substantially depending on the structure and optimization tasks of the DNNs \cite{Greta}. 

\begin{figure}[t!]        
\centering
    \includegraphics[width=1\linewidth]{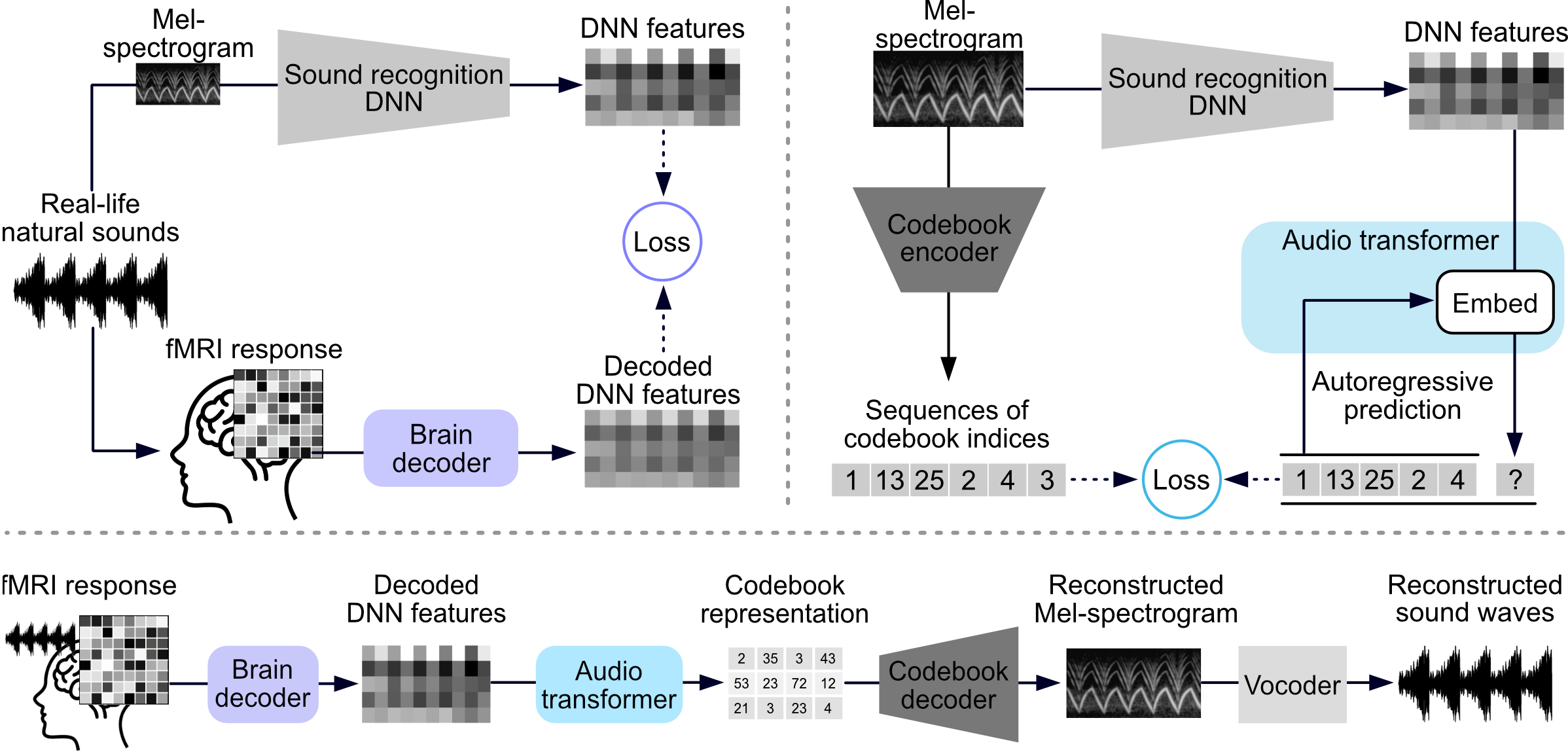}
\caption{{\bf Schematic overview of the proposed sound reconstruction model from brain activity. }
(A) Brain decoder training. Auditory features derived from real-world natural sounds are processed through various computational models. fMRI activity was recorded while subjects listened to natural sounds. fMRI signals were then used to train brain decoders to predict the values of corresponding auditory features. Among auditory features, the DNN features extracted from a sound recognition DNN model, which demonstrated superior decoding performance, were identified as brain-like features. (B) Audio transformer training. Codebook indices, which provide a concise representation of the Mel-spectrogram, are derived using a codebook encoder. An audio-transformer is trained to predict sequences of these codebook indices, conditioned on the DNN features, in an autoregressive manner. (C) Sound reconstruction from fMRI response. The reconstruction process begins with the computation of decoded DNN features from the fMRI response using the feature decoders. These decoded DNN features are subsequently transformed into codebook indices through the audio transformer. In the final steps, the codebook decoder and spectrogram vocoder transform these codebook indices into Mel-spectrograms and into sound waves.}
\label{fig1}
\end{figure}

Building on these insights, we present a novel approach to sound reconstruction from fMRI activity that is grounded in the hierarchical processing of auditory features akin to the human brain. Our methodology synergistically combines the decoding of auditory features with an audio-generative model, enabling the disentanglement of temporally compressed information within DNN features. 
 
Our initial step involves identifying "brain-like features" through a brain decoding analysis of auditory features. In this approach, we train a brain decoder to predict the auditory features of the presented sound stimuli from fMRI response patterns (Figure 1A). These auditory features include the pixels of a Mel-spectrogram, the modulation features, and the features from a sound recognition DNN model. Reinforcing previous findings, we identify the hierarchical features from a sound recognition DNN model as the most "brain-like," as they showed better decoding performance from brain activity compared to other auditory features. Then, we train an audio-generative transformer, a sequence-to-sequence model, to predict the codebook representation (a concise representation of a Mel-spectrogram) conditioned on the DNN features in an autoregressive manner (Figure 1B). In the test phase, we obtain decoded DNN features from fMRI responses using the feature decoders, and then transform them into codebook representations using the audio-generative transformer \cite{Iashin2021Taming}. The next step involves converting these codebook representations into spectrograms with the help of a codebook decoder. Finally, we transform these spectrograms into audio waveforms using a spectrogram vocoder (Figure 1C).
 
We evaluate this method by training and testing the model on a diverse set of natural sounds and the fMRI responses. We compare the results across the conditions where different sets of model components, brain regions, sound features, and training category sets were used. To test whether the reconstructed sounds reflect subjective perceptual experiences, we use fMRI responses obtained from the "cocktail party" condition, in which the subject selectively attends to a single sound in the presence of multiple sounds. Our results suggest that the model can reconstruct arbitrary sounds from multiple brain areas beyond the category domain of model training, capturing the rough content and quality of the original sound and the subjective perception.

\section*{Results}
We designed two fMRI experiments to investigate the neural responses elicited by natural sound stimuli and auditory selective attention. Five healthy subjects participated in the study, during which whole brain fMRI responses were collected as they listened to 8-s snippets of real-world sounds (see Materials \& methods: "Experimental design"). In the natural sound experiment, we presented 1,200 real-world sound stimuli for the training dataset and 50 sound stimuli for the test dataset (Figure 2). For the training dataset, we aimed to emulate the diversity of natural auditory environments. Consequently, each audio clip in the training dataset included a blend of different sound categories. For the test dataset, we selected four representative sound categories following procedures outlined in earlier studies \cite{Norman-Haignere2015Distinct}: Human speech, animal sounds, musical instruments, and environmental sounds. Our test dataset comprised 50 audio clips, with each clip exclusively representing one sound category (see "Materials \& methods: Stimuli"). Subjects were instructed to maintain their focus on the presented sound stimuli while performing a one-back repetition detection task. In the auditory selective attention experiment, we presented 24 pairs of superimposed sounds derived from the test dataset under diotic listening conditions, where the identical audio signal is delivered to both ears of the listener. Subjects were instructed to selectively attend to one of the stimuli based on a visual cue (see "Materials \& methods: Experimental design"). Data collection involved four repetitions of the training dataset. To increase the amount of available data, we extracted three cropped 4-s fMRI samples from each 8-s fMRI trial, with a 2-s overlap. This process resulted in a total of 14,400 training samples. For the test and attention trials, we collected data from eight repetitions. To improve the signal-to-noise ratio of the fMRI signals, we averaged the fMRI samples corresponding to the same stimuli (attention task) over eight repetitions. Consequently, we obtained 150 test samples with natural sound stimuli and an additional 144 test samples for the auditory selective attention dataset.

\begin{figure}[t!]        
\centering
    \includegraphics[width=.8\linewidth]{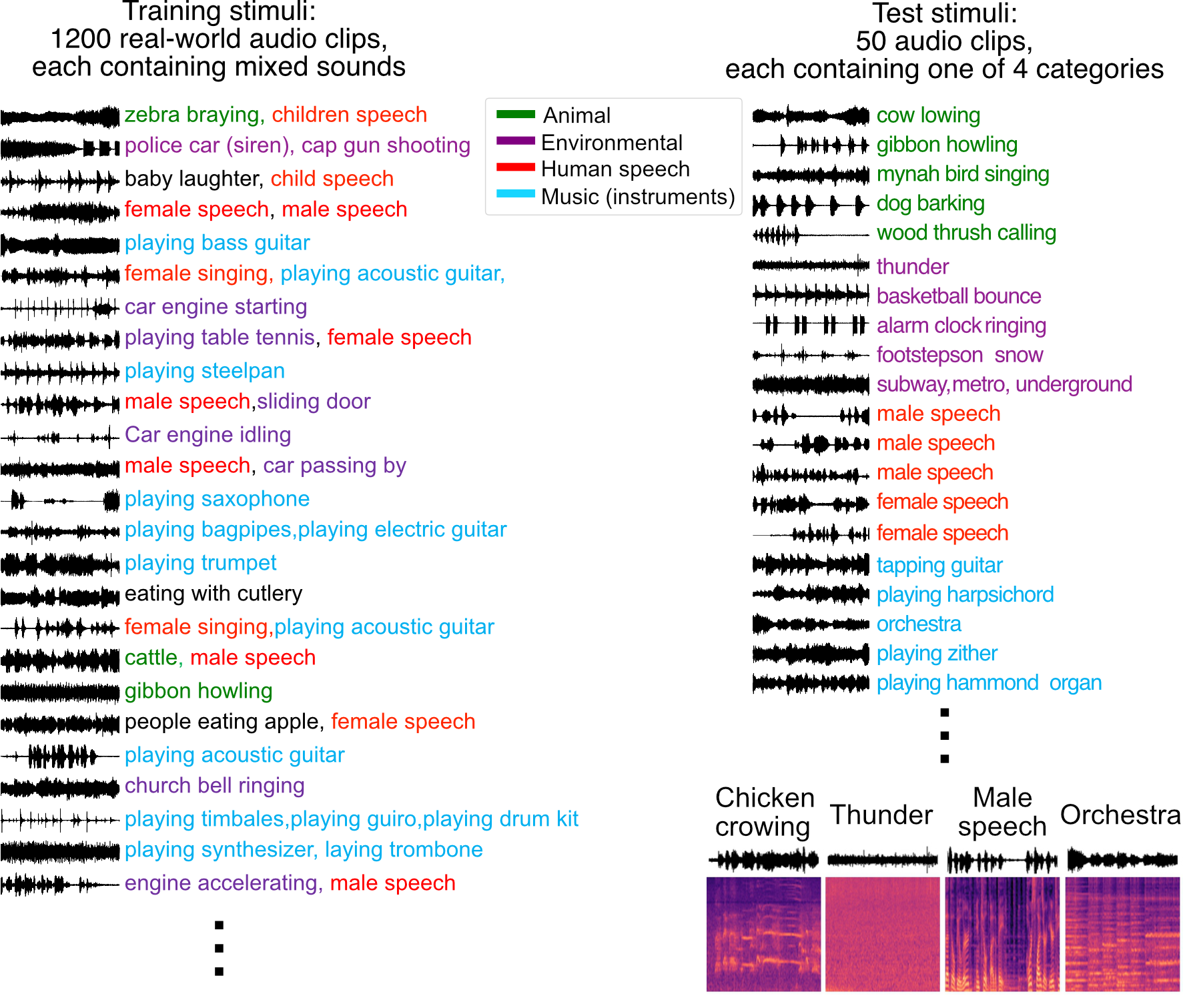}
\caption{{\bf Sound stimuli for fMRI Experiments.  }
The training dataset is comprised of 1,200 real-world audio clips (left panel). Each clip in this training set includes a mixture of sound categories, representing a broad array of natural auditory scenes. The test dataset consists of 50 audio clips, each containing only one sound category (right panel). The test categories include human speech, animal sounds, musical instruments, and environmental sounds, aligning with the categorization used in previous studies.}
\label{fig2}
\end{figure}

\subsection*{Brain decoding of auditory features}
We used L2-regularized linear regression to predict auditory features from the responses of thirteen anatomically defined ROIs located within the early auditory cortex and the auditory association cortex delineated from the Human Connectome Project (HCP) \cite{Glasser2016multi-modal}. We primarily visualized results from A1, LBelt, and PBelt in the early auditory cortex, as well as A4 and A5 in the auditory association cortex, following the ventral pathway. We also incorporated the responses from the combined all thirteen ROIs within early auditory and auditory association cortices to delineate an auditory cortex (AC) (Figure 3A). Using training fMRI samples of natural sound, we trained the decoder to predict auditory features: 1) pixel values of the Mel-spectrogram, 2) spectro-temporal modulation features, and 3) DNN features from the sound recognition model, VGGish-ish (Figure 3B). After the training phase, we used the brain decoder to predict the decoded feature values from the fMRI responses in the test dataset.
 
To evaluate decoding performance, we calculated the correlation coefficients between the decoded features from fMRI responses and the actual auditory features (Figure 3C). We found positive correlations across all combinations of feature types and ROIs from all subjects including our pilot study subject S1 (Figure 3D, Supplementary Figure 1: individual ROIs with hemisphere separation). Taking a closer look, most auditory ROIs demonstrated a correlation higher than 0.5 for the Mel-spectrogram and DNN features and showed a correlation above 0.4 for modulation features. Notably, the decoding performance for the Mel-spectrogram and modulation features progressively decreased as we moved from the A1 to the nonprimary cortex. In contrast, the DNN features retained correlations of above 0.5 up to A5 without a substantial drop in decoding performance. This suggests that while spectro-temporal features such as spectrograms and modulation features are effectively decoded from the early auditory cortex, their performance begins to wane as we move toward the peripheral regions. In contrast, the DNN features, although displaying decoding performance similar to the spectro-temporal features in the early auditory cortex, contain higher-level information through hierarchical processing akin to human auditory system. This attribute allows DNN features to be decoded even in the nonprimary auditory cortex. These findings collectively indicates the effectiveness of our trained brain decoder in predicting auditory features from fMRI signals.

Regarding the hierarchical combinations of DNN layers and individual ROIs' decoding performance (Supplementary Figure 2), we observed that early auditory cortical areas, such as A1, demonstrated the highest decoding accuracy compared to other auditory ROIs across most layers. In contrast, areas in the auditory association cortex, such as A4 and A5, exhibited slightly lower performance in the lower layers compared to A1, but their performance became comparable to A1 in the higher layers. This suggests that auditory cortical ROIs undertake distributed processing rather than operating as specialized areas with hierarchical homology, as mapped from the sound model.
 
We further evaluated the identification performance of the decoded features to determine their capability to identify perceived sounds among all test sounds. The identification accuracy was assessed by comparing the correlation coefficients between the decoded features and the actual auditory features of the all the test stimuli. This involved comparing the correlation between the decoded features and each candidate stimulus with the correlation between the decoded features and the actual stimuli that were presented. The identification accuracy of each decoded feature was quantified by the number of correctly identified pairs. Figure 3E indicates that the identification accuracy using decoded Mel-spectrograms for all subjects slightly exceeded the chance level in any auditory ROIs. Meanwhile, all subjects were capable of accurately identifying sound using decoded modulation features and decoded DNN features in the AC and individual ROIs. Notably, DNN features consistently outperformed other auditory features across all ROIs, with each subject achieving above 80\% identification accuracy. Our feature decoding analysis revealed that features derived from the hierarchical sound recognition model demonstrated superior predictive performance compared to the Mel-spectrogram or modulation features. As a result, we identified DNN features as the most "brain-like" features due to their enhanced decoding capabilities.
 
In the case of Mel-spectrograms, we encountered a peculiar discrepancy: the correlation coefficients of individual pixels across various stimuli are consistently high, yet the identification accuracy between the actual and predicted Mel-spectrograms is low. This anomaly may be attributable to the decoded Mel-spectrograms from the fMRI response, which appear to capture common variations across pixels rather than accurately decoding the actual values in each pixel. This empirical issue parallels previous findings reported in studies using direct regression to spectrogram features from neuroimaging techniques. In these studies, decoded Mel-spectrograms appeared to be dominated by an indistinguishable broadband component, leading to a smoothed pattern across the pixels \cite{Défossez2022Decoding}.

\begin{figure}[h!]            
\centering
    \includegraphics[width=.8\linewidth]{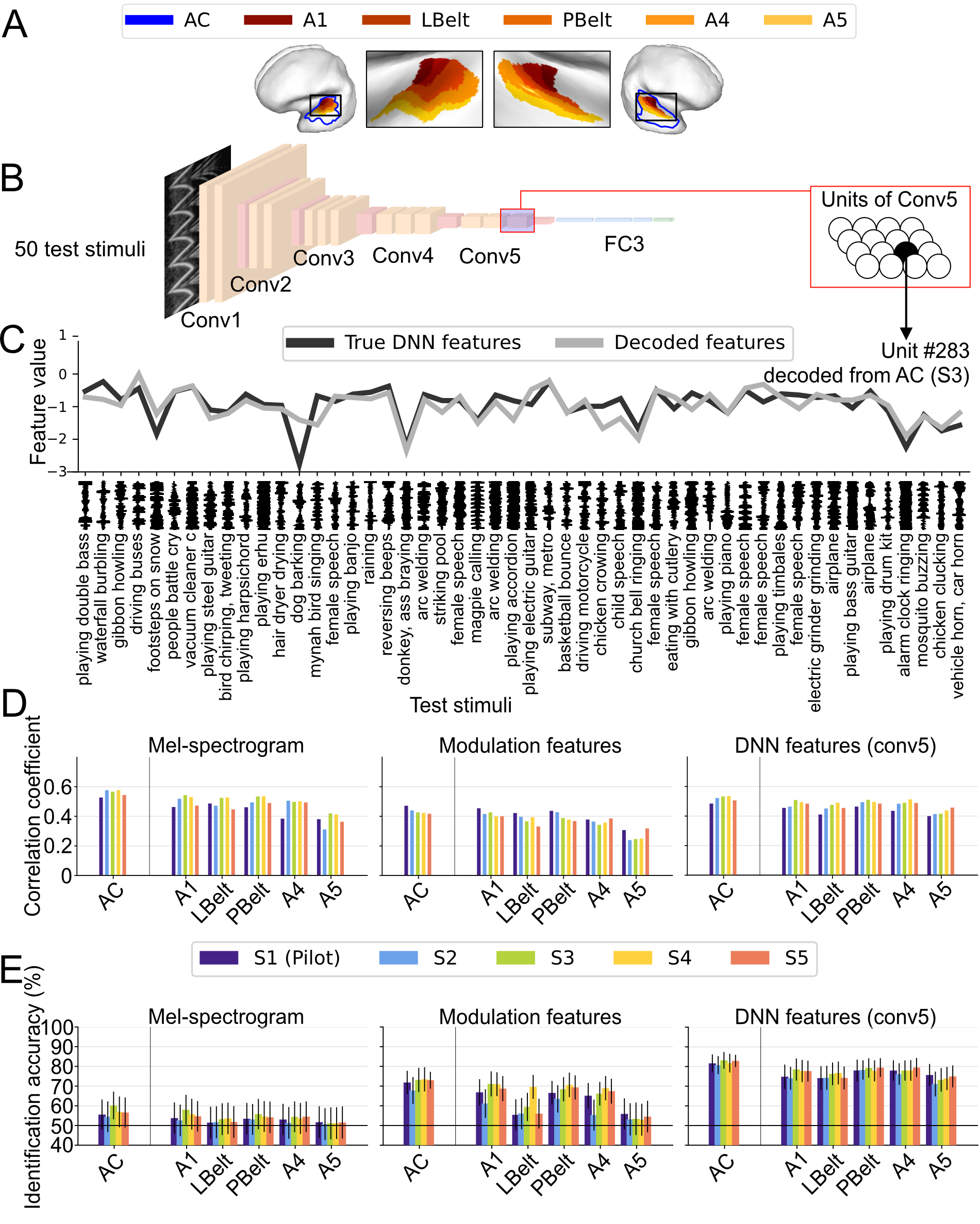}
\caption{{\bf Feature decoding analysis.  }(A) Region of Interest (ROI) definition. The auditory cortex (AC) is delineated as a combination of the early auditory cortex and auditory association cortex, based on the Human Connectome Project (HCP) parcellation. We visualized the A1, LBelt, and PBelt regions within the early auditory cortex, as well as A4 and A5 within the auditory association cortex, tracing along the ventral pathway. (B) Hierarchical structure of the Sound Recognition Model. The VGGish-ish model, designed for general-purpose sound recognition, is used in this analysis. The features computed from the highest convolutional layer (conv5) were used as the DNN features. (C) Example of true and decoded features for a DNN feature unit. This shows a comparison between the true DNN feature responses and those decoded from the AC for 50 test sound stimuli. This unit derived from the conv5 layer of the VGGish-ish model. (D) Comparison of decoding performance for three auditory features (Mel-spectrogram, modulation features, DNN features). Each bar corresponds to the average decoding accuracy of an individual subject, calculated across all units for each feature type. (E) Comparison of identification accuracy for three auditory feature types. To account for the lack of independence among partially overlapping samples, the identification accuracies of the 150 test samples were transformed into 50 data points by averaging the results from the three fMRI samples for each stimulus. Each bar represents the mean identification accuracy among these 50 data points, with an error bar indicating the 95\% confidence interval (CI). Each subject is represented by a different color.}
\label{fig3}
\end{figure}

\subsection*{Sound reconstruction}
In this section, we present our sound reconstruction process and evaluate the fidelity of reconstructed sounds from brain responses. From the feature decoding analysis, we identified the "brain-like" features which were extracted through the DNN model, mirroring the hierarchical structure of the auditory system. These DNN features encapsulate compressed temporal information. To convert these features back into their high-dimensional sound form, it is necessary to disentangle the compressed temporal information in DNN features. To accomplish this, we leveraged a transformer model notable for its remarkable performance in sequential processing. We trained an audio transformer to convert the sequence of DNN features into a sequence of codebook representations in an autoregressive manner. Using this trained transformer, we converted the decoded DNN features from fMRI responses into codebook representations. Subsequently, these codebook representations were transformed into Mel-spectrograms with a codebook decoder and then into audio waveforms using a spectrogram vocoder.  (see "Materials \& methods: Model components" and "Reconstruction methods").
 
Figure 4A provides examples of the reconstructed Mel-spectrograms from our model (see Supplementary Movies 1 for reconstructed sounds). The results indicate that the reconstructed spectral and temporal patterns that resemble the actual stimuli, highlighting the model's ability to reconstruct the auditory experiences from fMRI responses. Moreover, the reconstructed sounds from each participant demonstrated high reproducibility across different participants, indicating our approach's strong replicability.
 
Taking a closer look, the first four stimuli of Figure 4A showcase the reconstructed sounds for the animal category, where unique spectral patterns are reconstructed, rendering the reconstructed sounds easily identifiable as the corresponding actual stimuli. The subsequent four stimuli, representing the speech category, exhibit a clear harmonic pattern resembling human speech, distinguishing it from other categories. The following four stimuli, designated for the music category, display complex patterns spanning a wide frequency range, reconstructed in the Mel-spectrograms. The final four stimuli, assigned to the environmental category, present monotonous yet distinctive spectral patterns in the reconstructed Mel-spectrogram mirroring the actual sounds. Temporally rhythmic patterns, such as the bouncing of a basketball, are also reflected in the reconstructed sounds. Notably, unlike previous fMRI-based reconstructions, which presented temporally smoothed patterns, our model managed to reconstruct complex spectro-temporal patterns for each sound stimulus, preserving the rough contents, even though detailed content might not yet have been fully captured. 
 
To evaluate the fidelity and quality of the reconstructed sounds, we conducted a pairwise identification analysis. It assessed how accurately the reconstructed sounds could identify the stimulus between the pairs of stimuli where one is the true stimulus and the other is one of all the other test stimuli. To prepare for this identification analysis, we calculated auditory features from both the actual stimuli and the reconstructed sound, which incorporated pixels of the Mel-spectrogram, hierarchical auditory representation, and acoustic features. Figure 4B presents an example of a Mel-spectrogram used in the evaluation of the reconstructed sounds and acoustic features, specifically Fundamental Frequency (F0), Spectral Centroid (SC), and Harmonic to noise ratio (HNR). For each reconstructed sound, we calculated the correlation coefficient between the auditory features of the reconstructed sounds and those of the test stimuli. The identification accuracy for each reconstructed sound was then determined by counting the number of pairs where the actual stimulus was correctly identified among the candidates in the test set (see "Materials \& methods: Evaluation of reconstructed sounds"). 	
We first assessed identification accuracy by using the pixel values of the Mel-spectrogram (Figure 4C, left panel). All subjects demonstrated a mean identification accuracy across test stimuli of approximately 70\%. These results indicate that the reconstructions retain the raw-level feature information to some extent.
 
To assess the fidelity of the reconstructed sounds in hierarchical auditory features, we used the Melception classifier, a hierarchical sound recognition DNN model. Notably, the Melception classifier was trained independently, separate from the VGGish-ish model used for our decoding analysis. The extracted DNN features from the Melception classifier served as a surrogate for hierarchical auditory representations. The identification accuracy of reconstructed sounds, consistently across all subjects, exhibited superior performance when compared to the pixel-based analysis of the Mel-spectrogram, and the identification accuracy further increased with the progression of the hierarchical representation (Figure 4C, middle panel). This was particularly notable in the higher-level representations, where the mean identification accuracy across all test stimuli exceeded 85\% for each subject.
 
Despite our reconstruction being derived from higher-level DNN features, we sought to ensure that our reconstructed sounds represent the perceptual and qualitative aspects of the original. To confirm this, we evaluated the reconstructed sounds based on three acoustic properties: 1) F0, which gauges the perceived pitch and tonality, thereby distinguishing between different sounds and voices. 2) SC, which quantifies the brightness of a sound, with a higher spectral centroid usually indicating a brighter or sharper sound. 3) HNR, which differentiates between tonal sounds and noise-like sounds \cite{Alías2016Review}. When evaluated based on all these acoustic features, the mean identification accuracy for each subject was found to be comparable to the pixels of the Mel-spectrogram (Figure 4C, right panel). Specifically, the mean identification accuracy across test stimuli for each subject was approximately 70\% for both F0 and SC features and around 65\% for HNR. These results suggest that our proposed sound reconstruction model is capable of generating sounds preserving the rough perceptual quality characterized by the acoustic features to some extent.

To investigate the upper bounds of our model's sound reconstruction fidelity, we used actual features as input at each stage. Our proposed model produced a flawless reconstruction of the original (see Supplementary Figure 3). Using actual codebook representations as input led the codebook decoder to generate near-perfect reconstructions of the original sound (identification accuracy: 99\%). When we used actual DNN features as input for the proposed model instead of decoded DNN features, there was a slight dip in quantitative evaluation, though the results still demonstrated good fidelity (identification accuracy: 95\%). These results underscore that the fidelity of the reconstructed sounds are largely determined by the decoding performance of the brain decoding analysis.

We also conducted reconstruction analyses using decoded Mel-spectrogram features and modulation features. For Mel-spectrogram features, we predicted the pixel values of the Mel-spectrogram from fMRI responses and transformed these into a sound wave using a spectrogram vocoder. For modulation features, we employed a two-step iterative reconstruction algorithm which initially transitioned from modulation features to a spectrogram and then from the spectrogram to a sound wave, based on methods used in a previous study \cite{Santoro2017Reconstructing}. The sounds that were reconstructed from decoded Mel-spectrogram and modulation features appeared as temporally smoothed patterns of the original spectrogram (Supplementary Figure 4A and Supplementary Movie 3). A quantitative evaluation (Supplementary Figure 4B) showed that while the reconstructed sounds from the Mel-spectrogram and modulation features could identify the actual stimuli above-chance level, it only performed well in the mid-level hierarchical layers. On the other hand, our proposed model that employed DNN features outperformed other auditory features in all evaluated metrics.

\begin{figure}[!hbt]        
\centering
    \includegraphics[width=.8\linewidth]{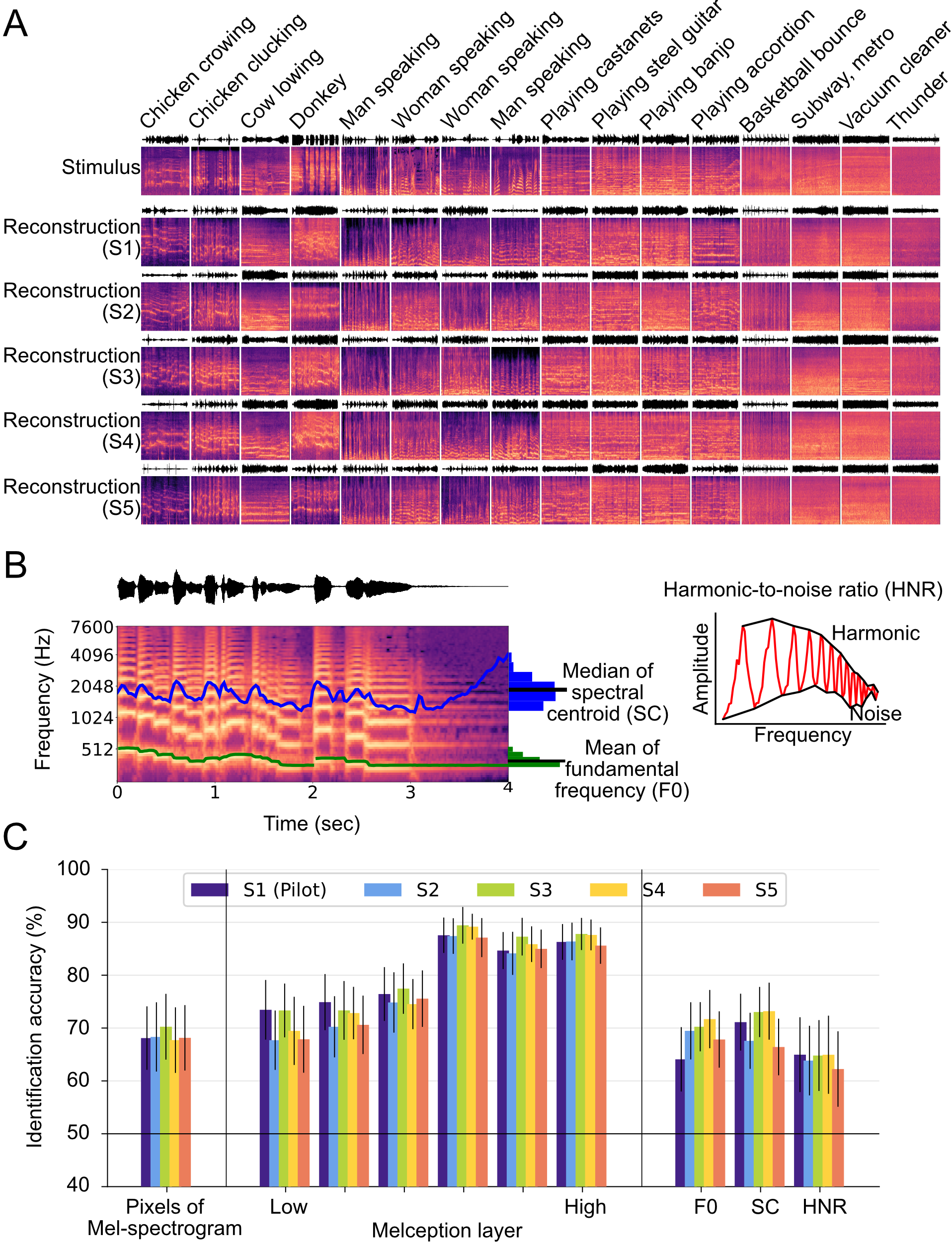}
\caption{{\bf Natural sound reconstruction.  }(A) Reconstructed Mel-spectrogram. The top row displays the original Mel-spectrogram of the presented sound. The subsequent five rows exhibit the Mel-spectrograms reconstructed from each subject using AC and conv5 layer. (B) Acoustic properties. This panel displays an example of a Mel-spectrogram used to assess the reconstructed sounds. Three key acoustic properties were evaluated - the Fundamental Frequency (F0), the Spectral Centroid (SC), and the Harmonic to noise ratio (HNR). (C) Evaluation of reconstructed sounds. The fidelity and quality of sound reconstruction was assessed via an identification analysis that incorporated pixels of the Mel-spectrogram, hierarchical representation, and acoustic features. Each bar represents the mean identification accuracy with an error bar indicating the 95\% confidence interval estimated with 50 data points. Each subject is represented by a different color.}
\label{fig4}
\end{figure}

\subsection*{Generalization beyond trained categories}
To confirm the capacity of our model to reconstruct beyond the categories of the training data, we conducted a post-hoc ablation analysis. In this analysis, the decoder was trained while excluding one category at a time from the training data, and the reconstruction results for the excluded category from the test dataset were evaluated. Supplementary Figure 5A demonstrates the reconstructed results indicating that even with an ablated training category set, the model maintains the reconstruction of spectral and temporal patterns resemble actual stimuli. Moreover, our model maintained reasonable reconstruction performance for the categories of animal and environmental sounds, even when these categories were excluded from decoder training (Supplementary Movie 3). For speech, despite the presence of substantial noise, human voices could still be distinctly discerned from the reconstructed sounds. However, when the music category was omitted during the training phase, there was a notable performance drop, resulting in reconstructed sounds resembling rhythmic environmental noises. In the quantitative assessment, we juxtaposed the identification accuracies from the ablated training category set with that from the full training category set. Despite a slight performance drop, the overall identification accuracies from the ablated training category set (Supplementary Figure 5B) were comparable to that of each category in full training category set (Supplementary Figure 5C). Looking at results by category, animal and environmental sounds showed above 70\% identification accuracy for most metrics in the ablated training set. The music and speech categories displayed performances around or below 60\% using pixels of Mel-spectrograms, F0, and HNR, and exhibited approximately 70\% accuracy at higher hierarchical representations. These results mirrored the tendencies observed in the full training set. The generalization ability even without the same category in the training dataset suggests that the proposed model is not simply matching brain data to the training examples. Instead, it appears to synthesize a sound through a combination of elemental features.

\begin{figure}[!hbt]        
\centering
    \includegraphics[width=.8\linewidth]{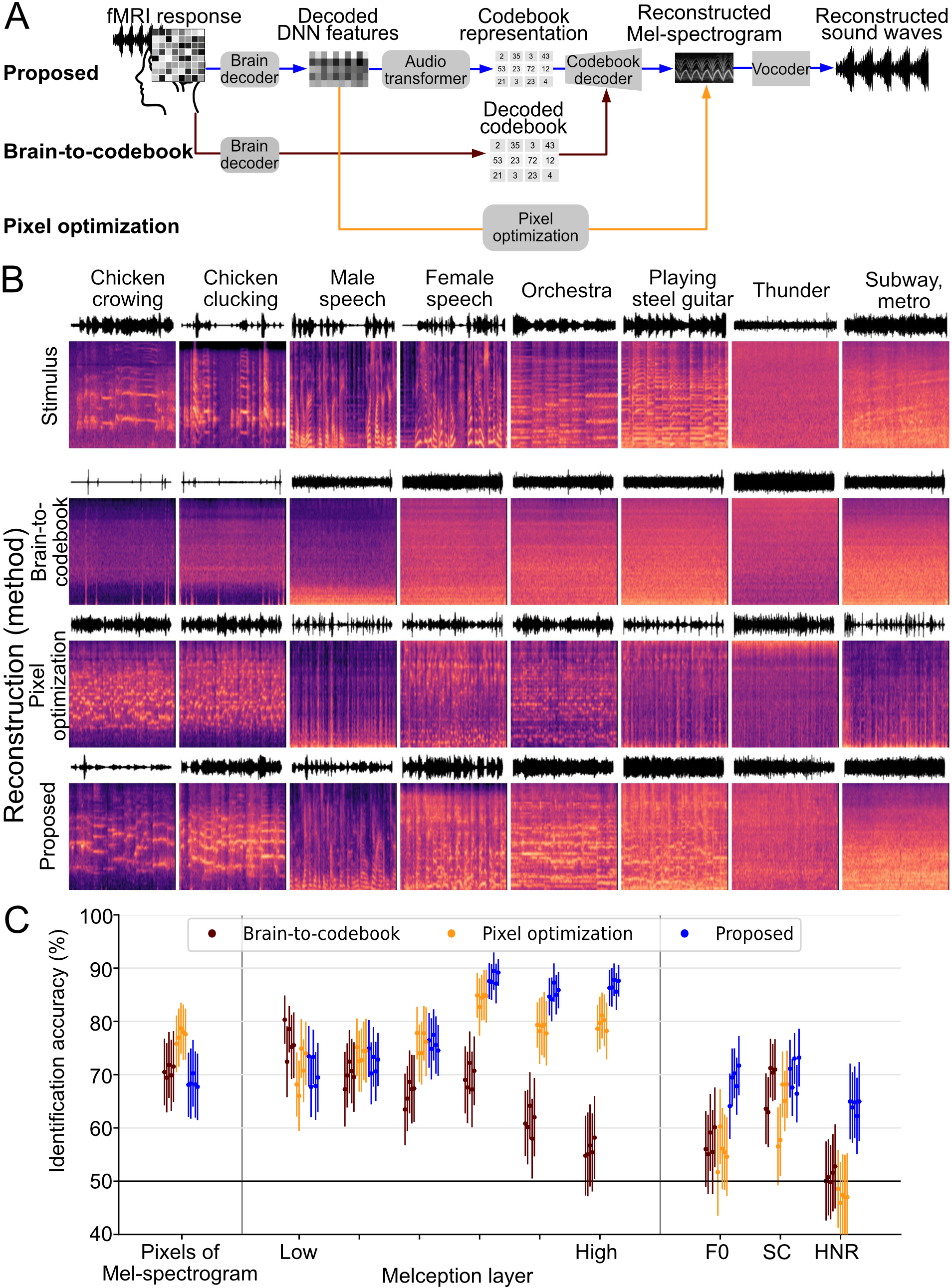}
\caption{{\bf Effect of model components.  }. (A) Overview of reconstruction methods. The brain-to-codebook reconstruction strategy predicts the codebook representation directly from the fMRI responses (depicted by the brown line). The pixel optimization reconstruction strategy iteratively optimizes the pixel values of Mel-spectrograms to align with the decoded DNN features inferred from fMRI responses (depicted by the orange line). (B) Reconstructed Mel-spectrograms. The original Mel-spectrogram of the presented sound is displayed in the first row. The subsequent rows two to four show the reconstructed Mel-spectrograms using different reconstruction methods from S3 using AC. (C) Evaluation of reconstructed sounds using different methods. Each method is represented by a unique color. Each dot represents the mean identification accuracy calculated from each subject, with the error bar indicating the 95\% CI estimated using 50 data points.}
\label{fig5}
\end{figure}

\subsection*{Model components}
To investigate the impact of different components within our proposed model, we compared the reconstructed results against two alternative reconstruction methods that used either decoded codebook representation or decoded DNN features (Figure 5A). The first approach, labeled as the brain-to-codebook reconstruction, employed a linear regression model to predict the codebook representation derived from the codebook encoder. The second method termed the pixel optimization reconstruction, iteratively optimized pixel values of the Mel-spectrogram to align with DNN features predicted from brain responses \cite{Shen2019Deep}.

In the brain-to-codebook reconstruction, the brain decoder was trained to predict codebook representations from fMRI responses, and we applied it to the test dataset. The decoding performance is displayed in Supplementary Figure 6. While it was inferior compared to DNN features, all auditory ROIs demonstrated a positive correlation. The decoded codebook representations were subsequently converted into a Mel-spectrogram using a codebook decoder and into sound waves with a spectrogram vocoder. The brain-to-codebook reconstruction generated a Mel-spectrogram that represented a temporally smoothed version of the original spectrogram, echoing previous discoveries on direct regression on modulations or physical features \cite{Santoro2017Reconstructing,Défossez2022Decoding} (see Figure 5B, Supplementary Figure 4A and Supplementary Movie 4). In the quantitative evaluation of the brain-to-codebook reconstruction (Figure 5C), all subjects showed identification accuracies, based on the pixels from the Mel-spectrogram, that were comparable to the proposed model, approximately 70\%. However, as the representation progressed towards higher layers within the hierarchical sound representation, identification performance progressively declined. At the highest Melception layer, all subjects exhibited performance lower than 60\%. In the evaluation using acoustic features, while SC showed comparable identification accuracy to the pixels of Mel-spectrogram, all subjects demonstrated identification performance below 60\% in both F0 and HNR metrics.
 
On the other hand, the pixel optimization reconstruction method leverages all decoded DNN features extracted from the VGGish-ish model to optimize the spectrogram pixels for sound reconstruction. While this approach reconstructs distinct spectral patterns, it fails to capture the detailed temporal patterns inherent in the original spectrogram (as shown in Figure 5B and Supplementary Movie 4). In the quantitative evaluation, the pixel optimization reconstruction demonstrates an identification performance based on the pixels of the Mel-spectrogram that is comparable to other reconstruction methods. However, when evaluated based on hierarchical representations, this method performed better than the brain-to-codebook reconstruction methods, but it still falls short of our proposed method, particularly in the higher layers. Furthermore, in the evaluation using acoustic features, the identification accuracy using F0 and HNR metrics was merely at the chance level, except for SC. This implies that while all three reconstruction methods managed to reconstruct the approximate spectral patterns to some extent, only reconstructed sound from the proposed model retained perceptual qualities akin to actual stimuli.
 
In summation, our proposed model outperformed other reconstruction methods in generating sound reconstructions that were more discernible and perceptually akin to the original sounds. This underscores the importance of integrating brain-like features and segregating temporal information from DNN features in our sound reconstruction approach.

\begin{figure}[!ht]        
\centering
    \includegraphics[width=.8\linewidth]{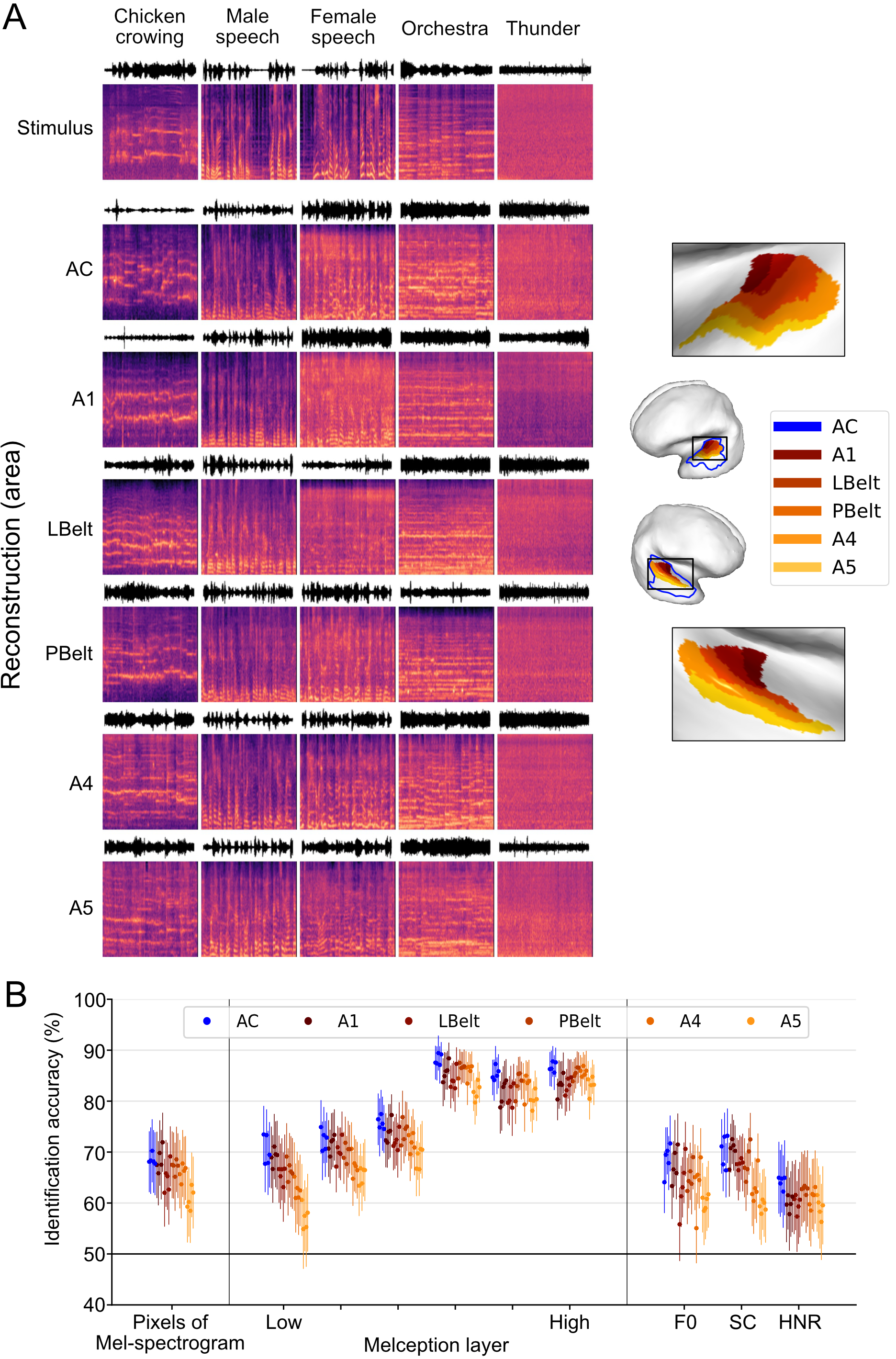}
\caption{{\bf Effect of auditory regions.   }(A) Reconstructed Mel-spectrogram from individual ROIs. The first row displays the original Mel-spectrogram of the presented sound. The subsequent rows, two through seven, depict the Mel-spectrograms reconstructed from each individual ROI using from S3 using conv5. (B) Evaluation of reconstructed sounds from individual ROIs. Each ROI is symbolized by a unique color. Each dot represents the mean identification accuracy calculated from each subject, with the error bar indicating the 95\% CI estimated using 50 data points.}
\label{fig6}
\end{figure}

\subsection*{Auditory ROIs}
We next explored the variability in reconstructed sounds across individual auditory regions. To facilitate this analysis, we trained the brain decoder exclusively on individual ROIs. Subsequently, we used the decoded DNN features obtained from each ROI to reconstruct the sounds. Figure 6A summarizes the reconstructed Mel-spectrogram from individual ROIs (refer to Supplementary Figure 7 for hemisphere separation). Our investigation found that the reconstructed sounds bore a resemblance to the original sounds, irrespective of the auditory ROI from which they were derived (Supplementary Movies 5). Additionally, the reconstructed sounds from each ROI demonstrated high reproducibility among themselves.
 
Figure 6B demonstrates the quantitative evaluation of ROIs. We further confirmed that the Mel-spectrogram and the low-level features within the hierarchical representation were more accurately reconstructed from the early auditory cortex, particularly region A1. Conversely, intermediate or high-level features within the hierarchical representation showed better reconstruction from the auditory association cortex, such as regions A4. The acoustic features showed higher identification performance when reconstructed from the early auditory cortex compared to the auditory association cortex. These findings underscore our proposed reconstruction model reflected the distributed neural responses within the auditory region.

\begin{figure}[!hbt]        
\centering
    \includegraphics[width=.8\linewidth]{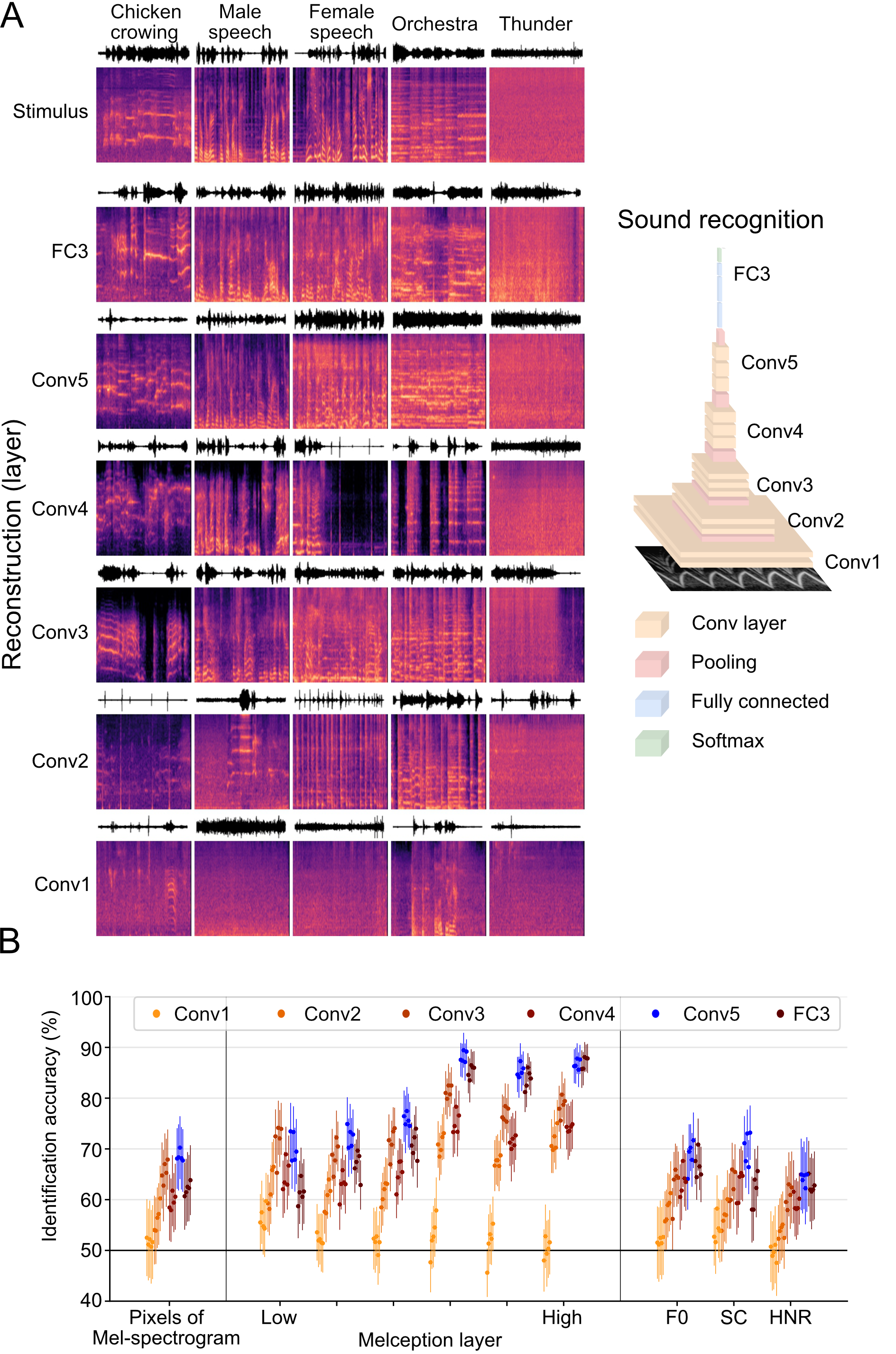}
\caption{{\bf Effect of DNN layers.   }(A) Reconstructed Mel-spectrogram using different DNN feature Layers. The original Mel-spectrogram of the presented sound is displayed in the first row. The subsequent rows, two to seven, represent the Mel-spectrograms reconstructed from different layers of the VGGish-ish model from S3 using AC. (B) Evaluation of reconstructed sounds from different VGGish-ish layers. Each layer is symbolized by a unique color. Each dot represents the mean identification accuracy calculated from each subject, with the error bar indicating the 95\% CI estimated using 50 data points.
}
\label{fig7}
\end{figure}

\subsection*{DNN layers}
We investigated the impact of the hierarchical representations of the DNN model on sound reconstruction. We trained transformers to predict codebook indices from each layer's DNN features and carried out reconstruction experiments. The results are summarized in Figure 7A and Supplementary Movie 6 (reconstructed from AC). Our findings revealed that lower layers, such as Conv1 and Conv2, produced reconstructions of low perceptual quality, characterized by considerable noise. On the other hand, intermediate layers, such as Conv3 and Conv4, exhibited spectral patterns similar to the actual stimuli. The sounds reconstructed from higher layers, particularly Conv5, mirrored the spectral patterns of the actual stimuli. Despite its lack of a temporal dimension, the FC3 layer was still capable of replicating distinct spectral patterns of the actual stimuli in the reconstructed Mel-spectrogram. We found that the sounds reconstructed from the higher layers retained perceptual content better than those reconstructed from the lower layers.
 
Figure 7B displays the quantitative evaluation of DNN layers. We found that sound reconstructions generated from higher layers showed enhanced identification accuracy than lower layers across all metrics, with the top convolution layer (Conv5) achieving the highest performance. On the contrary, sounds reconstructed from the FC3 layer showed comparable performance to Conv5 in high-level hierarchical representations but fell short in terms of Mel-spectrogram or lower-level representations. These findings indicate that incorporating DNN features with a temporal dimension enhances the low-level fidelity of the reconstruction.

\subsection*{Attention}
To ascertain that our reconstructed sounds encapsulate the subjective listening experiences, we collected fMRI responses under the selective auditory attention task known as "cocktail party conditions \cite{McDermott2009cocktail}." This situation illustrates the intricate capacity of our auditory system to isolate a specific sound from background noise in an environment with multiple sound sources. We aimed to verify if the decoded DNN features and the reconstructed sounds could reflect the process of selective attention. We derived decoded DNN features from the fMRI response during a selective auditory attention task, using a brain decoder trained under natural sound listening conditions. Using the decoded features, we produced sound reconstructions under the auditory attention tasks. 

Figure 8A summarizes the reconstructed sounds when the subjects were given the same stimuli of two superimposed sounds and attended to each sound respectively. The reconstructed sounds and Mel-spectrograms tend to resemble the attended stimulus more than the unattended one, even though they generally mirror the actual superimposed stimuli (Supplementary Movie 7). Particularly when attention is given to speech, the harmonic structure similar to the spectral pattern of speech is reflected in the reconstructed sounds.

To evaluate the capability of our reconstructed sounds to distinguish the attended stimulus from the unattended stimulus, we conducted an identification analysis based on the audio features extracted from the reconstructed sounds (Figure 8B). This analysis involved comparing the correlation between attended and unattended stimulus with the auditory features extracted from the reconstructed sound. Correct identification for each reconstructed sound was defined by its higher correlation with the attended stimulus than to the unattended one. The binary results from three samples of the same stimulus and attention task were pooled by majority voting to compute a single binary data point, leading to 48 binary data points. For each condition and subject, we computed the mean identification accuracy from 48 data points and compared it with the level of significance derived from the binomial test.

The identification performance was evaluated using the three types of extracted features as in the analysis of the single sound samples. As Figure 8C shows, when using Mel-spectrogram and low-layer features, most subjects' reconstructed sounds demonstrated an identification accuracy below 60\%. However, in the intermediate and high-layer features within the hierarchical representation, we observed an improved performance, with identification accuracy increasing to around 60\% for most subjects. In the intermediate and higher layers, some of the subjects demonstrated performance that exceeded the significance threshold. When evaluating the reconstructed sounds based on acoustic features, the identification accuracy for most subjects was less than 60\%, falling below the level of statistical significance.

In a subsequent analysis, we conducted a reconstruction analysis with individual auditory ROIs. Supplementary Figure 8A summarizes the reconstructed sounds under the selective attention task from individual ROIs. Regardless of the auditory ROI from which they were derived, the reconstructed sound from individual ROIs were generally similar (Supplementary Movie 8). However, spectral patterns resembling the attended stimuli, such as the harmonic spectral pattern of speech, were more apparent in the reconstructed sounds from boundary regions, particularly A4 and A5, than in the core regions. Supplementary Figure 8B summarizes the quantitative evaluation of the reconstructed sounds under the selective attention task across individual ROIs. When assessed with Mel-spectrograms and low hierarchical representations, no discernible difference in identification accuracy between auditory ROIs was observed. However, an evaluation based on intermediate and high hierarchical representations consistently revealed a gradual increase in identification performance from A1 to A5 in the reconstructed sounds from three subjects. On the other hand, evaluations based on acoustic features showed no substantial disparities between ROIs. These findings suggest that subjects may have paid more attention to the categorical aspects of the attended stimuli, with higher auditory regions being more closely associated with attentional modulation.

\begin{figure}[!hbt]        
\centering
    \includegraphics[width=.8\linewidth]{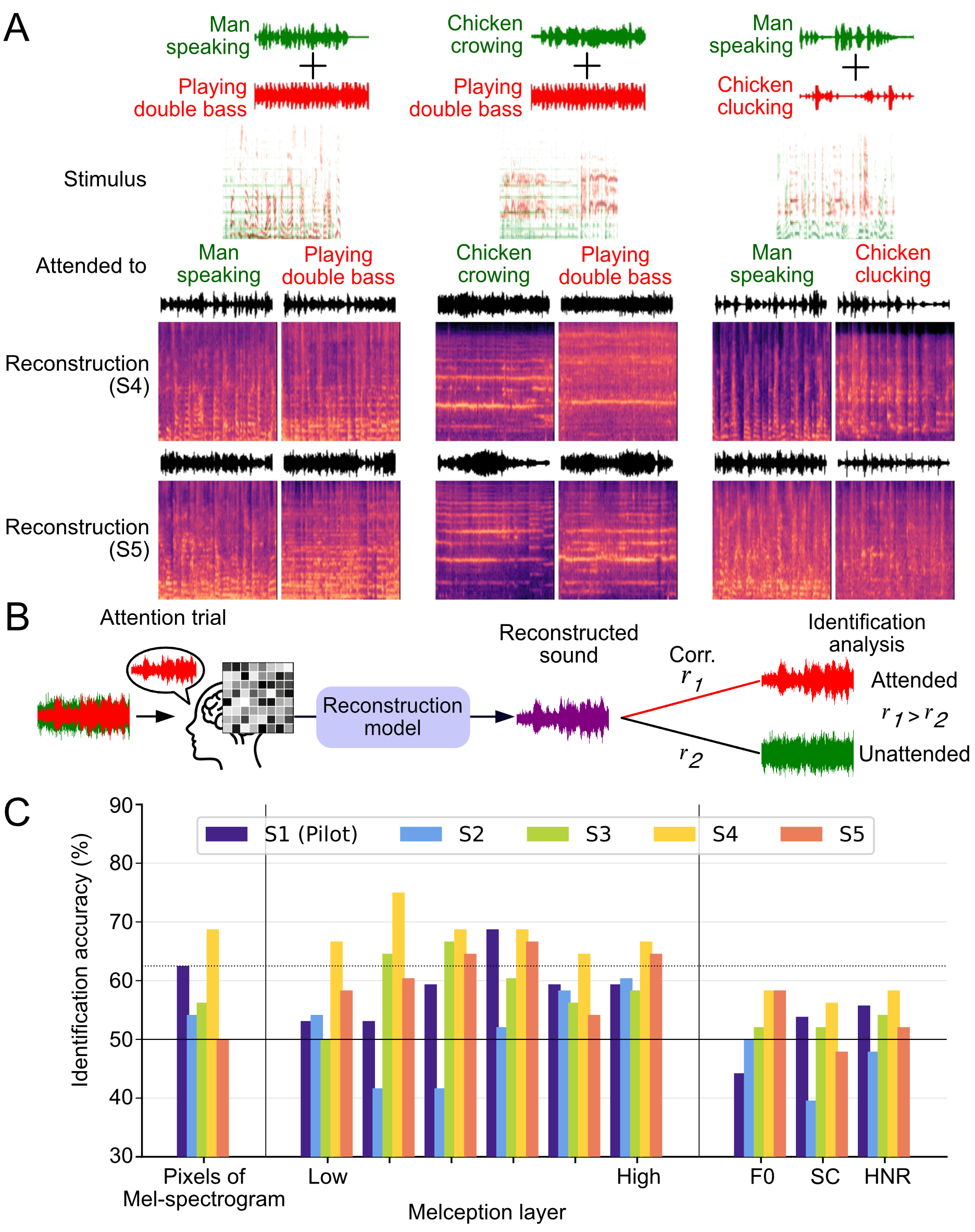}
\caption{{\bf Sound reconstruction of auditory selective attention.   } (A) Reconstructed Mel-spectrogram under selective auditory attention tasks. The top panel illustrates the Mel-spectrogram of the sound presented during the task. Here, subjects were instructed to focus on a specific sound from the superimposed sound stimuli. The bottom panel presents the Mel-spectrograms reconstructed from different subjects (S4, S5) using AC and conv5 layer. (B) Process for reconstruction and evaluation during selective auditory attention. The analysis involved evaluating the fidelity and quality of attention reconstruction through identification comparison between the attended and unattended stimuli. (C) Evaluation of sound reconstructed under selective attention tasks. The identification results within the three samples of the same stimulus and attention were pooled by major voting, resulting in 48 data points for each condition and subject. Each bar represents the mean identification accuracy of 48 data points. The dashed line represents the significant accuracy level (\textit{p} < 0.05) from the binomial test (\textit{N =} 48). Note that stimuli that failed in the calculation of F0 and HNR were excluded from the statistics. In such cases, a higher level of significant accuracy is required, though it's not depicted here.
}
\label{fig8}
\end{figure}
\section*{Discussion}
Our study presented a novel approach for unconstrained neural sound reconstruction from fMRI activity, leveraging "brain-like" auditory features and an audio-generative model. We conducted decoding analysis with a variety of auditory features from anatomically defined ROIs in the auditory cortex. From this analysis, we identified brain-like features derived from the sound recognition DNN model, mirroring the auditory system's hierarchical structure. To convert these decoded DNN features into high-fidelity audio signals, we used an audio-generative transformer to predict a concise representation of a Mel-spectrogram (a codebook representation) based on the decoded DNN features. Our model reconstructed complex spectral-temporal patterns, roughly maintaining the content and quality similar to the actual sound stimulus. This was validated using both qualitative and quantitative evaluations. Nevertheless, we acknowledged room for improvement, specifically in detailing the content within speech or music sequences. Importantly, our sound reconstruction remained robust even when certain categories were absent from the training phase. We extended our study to "cocktail party conditions," demonstrating the potential to reconstruct the subjective content of top-down auditory attention. The reconstructed sounds tended to reflect the attended stimulus over the unattended one.
 
In our analysis, we identified brain-like features that the features from sound recognition DNNs - which exhibit hierarchical processing akin to the human auditory system – consistently outperformed the decoding performance of other auditory features. These results align with previous encoding analyses of systematic model-brain correspondence \cite{Greta}. However, our decoding performance did not reflect a clear hierarchical correspondence between individual auditory ROIs and the layers within the DNN model, contrary to what previous encoding analyses with DNN had suggested \cite{Li2022Dissecting}. Recent studies using intracranial recordings have suggested a distributed functional organization within the human auditory cortex, pointing to a potential parallel information processing across the auditory cortex \cite{Hamilton2021Parallel,Nourski2014Functional}. Our results suggest that auditory cortical ROIs participate in distributed and hierarchical processing. During our decoding analysis employed anatomically defined ROIs, future studies may benefit from employing voxels specified by tonotopic or encoding analysis. This approach could provide further insight into the auditory hierarchy and its representations.

Although no discernable difference in feature decoding performance across individual ROIs was noted, we observed differences among individual ROIs when it came to the quality of the reconstructed sound. (Figure 6). We observed that the core region showed better identification performance in low-level representations and acoustic features, but the performance declined when moving toward the peripheral areas. In contrast, the high-level representations exhibited slightly better in the PBelt and A4 compared to A1. This finding aligns with previous encoding studies \cite{Norman-Haignere2022Multiscale}, suggesting that neurons in the vicinity of the primary auditory cortex engage in local integration and exhibit a selectivity towards generic acoustic representations. In contrast, neurons located near nonprimary regions partake in longer-time integration and show a preference for category selectivity. These findings suggest the potential of our proposed model to reconstruct the information encoded in individual ROIs.

While our model demonstrated proficiency in reconstructing sounds nearly perfectly using true features (Supplementary Figure 3), sounds reconstructed from decoded features lacked details in complex sequences such as speech and music. Recognizing these limitations, future investigations could benefit from exploring advanced decoding methods that incorporate sequential processing. Sequential techniques like Long Short-Term Memory networks (LSTM), Recurrent Neural Networks (RNN), and other recursive models have already seen extensive application in auditory decoding tasks using other neuroimaging modalities \cite{Daly2023Neural,Szabó2022Decoding,Yoo2021Decoding}. In addition, there have been efforts to exploit the inherent temporal information encapsulated within fMRI \cite{Li2019Interpretable,Loula2018Decoding,Wang2019Application}. Given our success in disentangling compressed features using a transformer model, employing a transformer to disentangle temporal information within fMRI signals may offer a potential pathway to enhance the quality and richness of the reconstructed sounds.

We expanded our reconstruction model to the cocktail party condition to verify whether our reconstruction reflected subjective listening experiences. Our results showed that the reconstructed sounds tended to resemble the attended sound than the unattended sound. This finding aligns with previous studies which have suggested that decoded auditory features, such as envelope \cite{Ding2012Emergence,O'Sullivan2015Attentional}, spectrogram \cite{O’Sullivan2017Neural}, and trajectory \cite{Bednar2020Where}, derived from brain responses during attention tasks, are more reflective of the attended sound rather than the unattended one. These studies commonly used neuroimaging techniques with high temporal resolution, such as MEG, EEG, and ECoG. Despite the limited temporal resolution inherent to fMRI, our study demonstrated the feasibility of reconstructing sound during attention-demanding tasks. Furthermore, our reconstructed sounds were more akin to the attended sound than the unattended one. These findings provide evidence that our reconstruction method mirrored the perceptual contents of subjective listening experiences under complex soundscapes.

In our analysis, we sought to improve the SNR of fMRI samples in both the test and attention experiments. In our analysis, we sought to improve the SNR of fMRI samples in both the test and attention experiments. Each fMRI sample was computed by averaging three consecutive functional volumes from the onset of the stimuli, as well as the trial average of the same test sound stimuli. Our pilot study demonstrated that these approaches improved both feature decoding performance and reconstruction quality. However, we also observed that the sounds generated from non-averaged fMRI samples still preserved the perceptual content and quality. Incorporating advanced decoding methods might offer potential paths for improving single-trial reconstructions. This has extensive implications for online reconstruction and the applications to brain–machine interfaces (BMI).

In our auditory attention analysis, we noted individual differences in reconstruction performance. In particular, subject S4, who substantially outperformed other subjects in feature decoding and sound reconstruction, reported an ease in focusing on a specific sound in a multi-speaker environment due to his extensive musical background. This suggests that personal experiences and task strategies may impact the decoding performance of selective attention. Therefore, it is advantageous to consider individual differences and task strategies in future research exploring the psychological and neural correlates of attention modulation.

Our proposed model introduces the possibility of reconstructing arbitrary sounds from neuroimaging responses by disentangling fine-grained temporal information from hierarchical sound features.
The potential of our method lies in its ability to externalize subjective auditory experience. By bridging the gap between internal cognitive processes and external manifestations, the reconstruction approach opens new ways for novel advancements in areas such as communication using imagined sounds, mental health diagnostics pertaining to auditory hallucinations, and artistic creation, as well as basic neuroscience research. Our proposed method contributes to the growing body of research that seeks to understand the neural basis of auditory perception and provides a valuable tool for understanding the human mind and fostering collaboration in the field of neuroscience.

\section*{Materials and methods}
\subsection*{Subjects}
Five non-native English-speaking subjects with normal hearing abilities, including one female subject, participated in our study. The average age of the subjects was 27.6 years. One subject (S1) was used for the exploratory analysis to establish the reconstruction model, while the other four subjects were used to validate the results independently. All subjects provided informed consent prior to the scanning sessions. The study protocol received approval from the Ethics Committee of the Advanced Telecommunications Research Institute International (approval no: 106) and was conducted following the principles of the Declaration of Helsinki.

\subsection*{Stimuli}
Our fMRI experiments for natural sounds used a total of 1,250 audio clips, each lasting 8-s, from the VGGsound test dataset \cite{Chen2020Vggsound:}. All stimuli were meticulously evaluated for their sound quality by human listeners. For the training dataset, we selected 1,200 audio clips indiscriminately, without considering category label information aiming to represent natural auditory scenes. Each audio clip in the training set might contain multiple sound categories. Any stimuli that caused categorization difficulties during the pilot study (S1) - totaling 162 - were replaced with new stimuli in the subsequent experiments with the four subjects. For the test dataset, we carefully selected four representative sound categories, following the criteria established in prior studies \cite{Norman-Haignere2015Distinct}]: Human speech (including English), animal, musical instrument, and environmental sound. We excluded categories not suitable for long sound stimuli, such as non-speech vocalizations. This resulted in a test dataset of 50 audio clips, each exclusively representing a single sound category. We resampled all the audio clips included in the fMRI dataset to 22050 Hz, center-cropped them to 8-s, and normalized them to ensure equal energy levels.

During the auditory selective attention experiment, we simultaneously presented pairs of sound categories, including male speech, female speech, animal sounds, and music. We excluded environmental sounds due to the challenges of focusing on these sounds when they overlapped with others. In the pilot study, we randomly selected two exemplars from each category. For the other four subjects, we chose two exemplars from each category based on those that demonstrated the highest reconstruction performance in the pilot study with subject S1. By combining all possible sound pairs, excluding pairs from the same category, we generated a set of 24 stimuli. In each attention trial, subjects were instructed to focus on one of the two concurrently presented sounds. Therefore, we obtained 48 attention trials for the attention dataset. The energy level of these superimposed sounds was normalized to match the level.

\subsection*{Experimental design}
Our study involved two fMRI experiments: the natural sound presentation experiment and the selective auditory attention experiment. During both experiments, we presented sound stimuli via fMRI-compatible headphones (Kiyohara, KAS-3000HK) at approximately 68-75 dB SPL. We adopted a continuous sampling approach, recording fMRI responses throughout the simultaneous presentation of stimuli without any silent interstimulus gaps. Subjects were allowed to adjust the sound level to their comfort before scanning.
 
In the natural sound presentation experiment, subjects passively listened to a variety of audio clips of natural sounds. We recorded whole-brain fMRI responses while subjects listened to 1,200 stimuli designated for training and 50 for the test dataset. Each subject underwent a series of scanning sessions spread over approximately three months, with 12-16 training sessions followed by a separate single test session. Every session included 4-8 functional runs, each not exceeding 90 minutes. Each run started with a rest period of 30-s, followed by 55 stimulus presentation blocks of 8-s each (comprising 50 unique sound stimuli and five randomly interspersed behavioral task blocks), and ended with a 10-s rest period. This sequence resulted in a total run duration of 8 minutes. To maintain subject concentration, we incorporated a one-back repetition detection task, where subjects were required to press a button if the subsequent stimulus presented was identical to the previous one. These repetition blocks (five per run) were not included in the analysis. 
 
In the selective auditory attention experiment, we instructed subjects to attend to one of two superimposed sound stimuli under diotic listening conditions. Each subject participated in a single session consisting of eight functional runs, with 48 attention trials in each run. In every attention trial, we simultaneously presented two different categories of sounds for 8-s. Along with the sound stimuli, we displayed visual cues that included words from both the attended and unattended sound categories. We distinctly marked the attended category word with a dash ("-") next to it, designating it as the target focus. Between trials, we randomly inserted a behavioral task four times in each run, where subjects were asked to identify the sound category they had focused on in the preceding trial.

\subsection*{MRI acquisition}
fMRI data were acquired using a 3.0-Tesla Siemens MAGNETOM Verio scanner at the Kyoto University Institute for the Future of Human Society. An interleaved T2*-weighted gradient-echo echo planar imaging (EPI) sequence was used to obtain functional images that covered the entire brain (TR = 2000 ms, TE = 44.8 ms, flip angle = 70 deg, FOV=192 $\times$ 192 mm, voxel size=2 $\times$ 2 $\times$ 2 mm, slice gap = 0 mm, number of slices = 76, multiband factor = 4). T1-weighted magnetization-prepared rapid acquisition gradient-echo (MP-RAGE) fine-structural images of the entire head were also obtained (TR = 2250 ms, TE = 3.06 ms, TI = 900 ms, flip angle = 9 deg, FOV = 256 $\times$ 256 mm, voxel size=1.0 $\times$ 1.0 $\times$ 1.0 mm, number of slices = 208).

\subsection*{MRI data preprocessing}
We used the same in-house MRI preprocessing pipeline based on fMRIprep that we outlined in our previous publication \cite{Horikawa2022Attention}. We adjusted the preprocessed functional data by shifting them forward 2-s to account for the hemodynamic delay. To increase the number of available data samples, we slid a 4-s time window across the original 8-s stimulus at 2-s intervals. For each of these 4-s sound stimuli, we created an fMRI sample by averaging the three consecutive functional volumes following the stimulus onset. This procedure resulted in three data samples from each original 8-s trial. As a result  , we obtained a total of 14,400 training samples (1,200 stimuli $\times$ 4 repetitions $\times$ 3 samples = 14,400 samples). For the test and attention datasets, we enhanced SNR by averaging the fMRI responses to identical sound stimuli across multiple repetitions. This approach yielded a total of 150 test samples (50 stimuli $\times$ 3 samples = 150 samples) and 144 attention samples (48 stimuli $\times$ 3 samples = 144 samples).

\subsection*{Regions of interest}
We used a multi-modal cortical parcellation developed by the Human Connectome Project (HCP) \cite{Glasser2016multi-modal} to delineate ROI within the auditory cortex. We identified thirteen anatomical ROIs in the early auditory cortex (A1, LBelt, MBelt, PBelt, RI) and auditory association cortex (A4, A5, TA2, STGa, STSd anterior, STSv anterior, STSd posterior, and STSv posterior) in both hemispheres. We considered the combined voxels of the early auditory cortex and auditory association cortex as the AC.

\subsection*{Model components}
In this study, we used multiple DNN models that were originally developed in the previous research conducted by Iashin and Rahtu \cite{Iashin2021Taming}. Pretrained models and scripts can be accessed through the following link: https://iashin.ai/SpecVQGAN. Specifically, we directly employed the pre-trained models for VGGish-ish, Melception classifier, and spectrogram vocoder, as they are independent of sound length. For tasks involving the generation of 4-s sound, in line with our fMRI experiments, we used the provided scripts for training models for Spectrogram Vector Quantized Generative Adversarial Network (SpecVQGAN) and audio transformer. 

We used the VGGsound dataset \cite{Chen2020Vggsound:}, a public repository of audio and video data extracted from YouTube. This dataset comprises 200,000 videos with reliably annotated labels spanning 309 categories. We excluded any videos with missing or lost links, yielding a balanced validation and test dataset, each with an equal number of audio clips across the 309 categories. The training dataset included 157,000 audio clips, while the validation and test datasets contained 19,000 and 15,000 audio clips, respectively. All audio clips were center-cropped to a duration of 4-s.

For data preprocessing, we resampled audio clips at a frequency of 22050 Hz and generated log-Mel-spectrograms using a short-term Fourier transform (STFT) with 1024 bins, 256 hop lengths, and 80 Mel band scales. These scales were centered on frequencies ranging from 125 to 7600 Hz. We further processed the resulting Mel-spectrograms by center-cropping in the time domain from  ($n_{spectral}$ $\times$ $n_{temporal}$ = 80 $\times$ 345) to ($n_{spectral}$ $\times$ $n_{temporal}$ = 80 $\times$ 336) to ensure compatibility with subsequent downscaling processes during the training phase.

\subsubsection*{VGGish-ish classifier}
The VGGish\-ish, a convolutional neural network (CNN) model composed of 13 convolution layers and three fully connected layers, was purposefully trained for sound recognition tasks using the VGGsound training dataset. This model was used to extract DNN features from Mel-spectrograms. The unit responses from each layer within the Mel-spectrograms were calculated as DNN features, with dimensions of ($n_{spectral}$ $\times$ $n_{temporal}$ $\times$ $n_{channels}$). These extracted DNN features were then reshaped to the format ($n_{spectral * channels}$ $\times$ $n_{temporal}$), preserving the temporal dimension, for use as conditioning input in the audio transformer model. We selected six representative layers that demonstrated superior decoding performance within each convolutional and fully-connected layer block: conv1\_1 with dimensions of (5120 $\times$ 336), conv2\_1 with dimensions of (5120 $\times$ 168), conv3\_1 with dimensions of (5120 $\times$ 84), conv4\_1 with dimensions of (5120 $\times$ 42), conv5\_3 with dimensions of (2560 $\times$ 21), and fc3 with dimension of (309).

\subsection*{SpecVQGAN}
We adopted a SpecVQGAN to achieve two objectives: compactly extracting discrete codebook representations and assuring the reconstruction of high-quality sound from these codebook representations. SpecVQGAN is a variant of the Vector Quantized Variational Autoencoder (VQVAE) model and incorporates vector quantization methods \cite{Walker2021Predicting,Razavi2019Generating,Oord2018Neural} to convert latent features into discrete units. This strategy aids in circumventing the issue of posterior collapse, a problem often arising when there is too much complexity in the model or insufficient constraints in the generative model's latent space. SpecVQGAN employs an encoder, which is a standard 2D-Conv stack with self-attention layers added that operate on an encoded representation. Our decoder mirrors the encoder's architecture, except for an upsampling layer that doubles the spatial resolution before the convolutional kernel with nearest-neighbor interpolation. By adhering to the default parameter from the referenced paper, the 4-s Mel-spectrogram in the shape of ($n_{spectral}$ $\times$ $n_{temporal}$ = 80 $\times$ 336) yielded concise codebook indices in the shape of ($n_{spectral}$ $\times$ $n_{temporal}$ = 5 $\times$ 21). 

\subsubsection*{Audio transformer}
For high-fidelity sound reconstruction from fMRI data, we applied a transformer model to disentangle compressed temporal information embedded within DNN features. Drawing inspiration from the demonstrated success of autoregressive generative applications \cite{Iashin2021Taming,Vaswani2017Attention,Esser2021Taming}, we trained an audio transformer to predict codebook representations for each spectral direction at every temporal point, consistent with an autoregressive approach. We used a GPT-2-medium transformer model that comprises 24 layers, 1024 hidden units, and 16 attention heads. This model's input consisted of DNN features with a shape of ($n_{spectral * channels}$ $\times$ $n_{temporal}$). The transformer processed the DNN features at each temporal point and transformed them into a probability distribution for the forthcoming codebook index. This conversion process was facilitated using a 1024-way softmax classifier. The training of the transformer was steered by the objective of minimizing the cross-entropy loss between predicted and actual codebook representations. As a result, the audio-transformer converts the sequence of DNN features into a sequence of codebook representations of shape ($n_{spectral}$ $\times$ $n_{temporal}$ = 5 $\times$ 21).

\subsubsection*{Spectrogram vocoder}
We employed MelGAN \cite{Kumar2019MelGAN:}, a generative adversarial network (GAN) for audio synthesis, to reconstruct audio waveforms from the Mel-spectrogram. Unlike the Griffin-Lim procedure \cite{Griffin1984Signal}, MelGAN generates audio waveforms in a non-autoregressive manner. This approach allows for a quicker and more high-fidelity audio reconstruction.

\subsection*{Auditory features}

\subsubsection*{Pixel values of Mel-spectrogram}
In our decoding analysis, we first used the pixel values of the Mel-spectrogram as an auditory feature. These Mel-spectrograms were calculated from all stimuli used in fMRI experiments, following the same data processing procedures as described in the model components. Each stimulus resulted in a Mel-spectrogram with a shape of ($n_{spectral}$ $\times$ $n_{temporal}$ = 80 $\times$ 336). 

\subsubsection*{Spectrotemporal modulation features}
We calculated spectrotemporal modulation features as another auditory feature for our decoding analysis, adhering closely to the methods outlined in \cite{Santoro2017Reconstructing}. First, this process involved generating audio spectrograms using a bank of 128 overlapping bandpass filters, evenly spaced along a logarithmic frequency axis. The output from this filter bank went through several processing stages, including bandpass filtering, frequency axis differentiation, half-wave rectification, and short-term temporal integration. We then computed the modulation content of the auditory spectrogram using a bank of 2D modulation-selective filters, performing a complex wavelet decomposition. This process resulted in a representation composed of $\times$ 20 temporal modulation frequencies $\times$ 10 time bins $\times$ 128 frequencies, totaling 153,600 features. Note that the audio spectrogram used for the modulation calculation is a time-frequency representation similar to, but different from, the Mel-spectrogram. However, to facilitate a comparison with earlier fMRI-based reconstructions, we opted to use the audio spectrogram rather than the Mel-spectrogram.

\subsubsection*{DNN features}
Lastly, we used DNN features derived from the Mel-spectrogram as an auditory feature. The previously computed Mel-spectrograms were processed through a pretrained VGGish-ish classifier, from which we recorded unit responses from its highest convolution layer, conv5\_3. This produced DNN features with a shape of ($n_{channels}$ $\times$ $n_{temporal}$ $\times$ $n_{temporal}$ = 512 $\times$ 5 $\times$ 21). To incorporate temporal sequences of these DNN features as inputs into the auditory transformer, these features were reshaped to preserve temporal information. Consequently, we used the DNN features in the shape of ($n_{spectral * channels}$  $\times$ $n_{temporal}$ = 2560 $\times$ 21). 

\subsection*{Feature decoding analysis}
We implemented a brain decoding framework to predict auditory features from multi-voxel fMRI responses, as outlined in our prior works \cite{Horikawa2022Attention,Horikawa2017Generic,Horikawa2020Neural}. For training the brain decoders, we employed 14,400 samples from the training dataset for each combination of auditory features and brain areas. The training process began with voxel selection, guided by the correlation coefficient with the target features. We selected 500 voxels from AC, and 200 from each individual ROI. These selected voxels constituted the input to the decoders. Next, we calculated the mean and standard deviation of the responses from the selected voxels in training samples, and used these values to normalize the voxel responses. We also applied z-score normalization to the feature values of the stimulus. This process used the means and standard deviations of the feature values estimated from the training data. Finally, we applied an L2-regularized linear regression model to predict normalized feature values based on the multi-voxel patterns of the normalized fMRI responses.

In the testing phase, each fMRI sample from the test dataset was first normalized using the mean and standard deviation obtained from the training dataset. The trained decoders were then applied to these normalized samples to predict the auditory features from 150 fMRI samples. These decoded features were subsequently denormalized using the mean and standard deviation for each feature derived from the training dataset. Given the potential discrepancies between the distribution of actual and decoded features, we implemented a posthoc normalization operation where the decoded feature values were normalized by the square root of the number of repetitions. During this process, we ensured that the mean of the posthoc normalized decoded features remained consistent with that of the decoded features calculated from brain decoder.

To assess decoding performance, we computed two metrics: 1) the Pearson correlation coefficient between the actual and decoded auditory features across the test stimuli in each pixel or unit and 2) an identification analysis to evaluate the ability of the decoded auditory features to identify the actual stimuli from a set of candidate stimuli.

For the auditory attention experience, we used brain decoders that had been trained under the passive listening conditions. These decoders predicted auditory features from the 144 fMRI samples in the attention dataset. The decoded auditory feature values were used as inputs to the reconstruction model. To assess the decoding performance, we conducted an identification analysis on the attended and the unattended stimuli based on the correlations with the decoded auditory features.

\subsection*{Reconstruction methods}

\subsubsection*{Proposed model}
Our reconstruction model, combining the trained model components and the brain decoder, operates in the following sequential manner. First, the brain decoder decodes DNN features from fMRI responses in the test dataset. These decoded DNN features are then transformed into codebook representations with the help of an audio transformer. These codebook representations are further converted into Mel-spectrograms using a codebook decoder. Ultimately, a spectrogram vocoder transforms these spectrograms into audio waveforms. Since our model produces various intermediate representations, we performed an ablation analysis to examine the influence of each model component during the reconstruction process. The brain-to-codebook reconstruction approach circumvents the use of brain-like features, directly predicting codebook representations from brain responses. Conversely, the pixel optimization reconstruction excludes the audio transformer, typically used to disentangle temporal information in decoded brain-like features, and instead directly optimizes the Mel-spectrogram from the decoded DNN features.

\subsubsection*{Brain-to-codebook reconstruction}
The process started with training a decoder to predict codebook representations directly from fMRI responses, and we computed decoded codebook representations from the test dataset. The next step involved transforming these decoded representations into quantized codebook representations by identifying the nearest representations in the pre-trained codebook dictionary. These quantized codebook representations were then converted into Mel-spectrograms using the codebook decoder. In the final stage, we transformed these Mel-spectrograms into audio waveforms using the spectrogram model.

\subsubsection*{Pixel optimization reconstruction}
We implemented an image feature-based optimization technique, optimizing pixel values in 2D Mel-spectrogram images using the VGGish-ish model \cite{Shen2019Deep}. Unlike techniques that optimize RGB images, our method was designed for grayscale Mel-spectrograms. We based our implementation on open-source code available at https://github.com/KamitaniLab/DeepImageReconstruction. The algorithm starts with a noisy initial image and iteratively optimizes pixel values to align the DNN features extracted from the VGGish-ish model with those decoded from brain activity across all DNN layers. Notably, we encountered a similar issue mentioned in the referenced paper: a lack of loss convergence when attempting reconstruction using only features from the single layer, specifically conv5\_3 of our proposed model. To overcome this, we minimized the loss across all features of the VGGish-ish layers in comparison with the decoded features. All other parameters were maintained at their default settings.

\subsection*{Evaluation of reconstructed sounds}
To evaluate the fidelity and quality of our reconstructed sounds, we carried out a pairwise identification analysis. During this procedure, we extracted auditory features from both the original stimuli and the reconstructed sounds. We then assessed the accuracy with which the reconstructed sounds could correctly identify the original stimulus, in pairs comprising one true stimulus and each of the remaining test stimuli.

Fidelity was evaluated using the pixels of the Mel-spectrogram and hierarchical representations. Initially, we assessed fidelity using Mel-spectrogram pixels, which are raw-level features. Following this, we employed a pre-trained Melception classifier—an audio classifier specifically trained for sound recognition tasks—to examine hierarchical representations' fidelity. We extracted DNN features from both the reconstructed sounds and the original stimuli using this classifier, which served as a surrogate for hierarchical sound representations. For this process, we utilized six representative layers of the Melception classifier (conv1, conv5, mix5\_d, mix6\_d, mix7\_c, and fc1).

To assess the quality of reconstructed sounds, we extracted three key acoustic features: 1) Fundamental frequency (F0), which represents the lowest frequency in a sound's periodic waveform and is associated with pitch perception. 2) Spectral centroid, which denotes the center of mass of the spectrum and relates to the balance of frequency amplitude, and it's closely linked with the perception of "brightness" of a sound, 3) Harmonic to noise ratio, which represents a measure of sound quality that compares the level of periodic, harmonic components of the sound to the level of noise sound. We segmented each sound stimulus into short sections and computed the F0 based on PYIN methods \cite{Mauch2014PYIN:}, and SC using the Librosa toolbox (https://librosa.org). Subsequently, the HNR was calculated using an implemented Python package (https://github.com/brookemosby/Speech\_Analysis). However, there were instances where we could not compute the F0 or HNR from each segment or the entire stimulus of both the true and reconstructed sounds, particularly when the sound contained a non-harmonic structure without a discernable pitch. In these cases, these stimuli were excluded from analysis. The mean value (or median for SC) from all segments represented the acoustic feature of each stimulus and reconstruction.

Using the extracted auditory features, we calculated the correlation coefficient between the auditory features of the reconstructed sounds and those of a pair of candidate stimuli. For the single sound test samples, one of the candidates was the presented stimulus, and the other was one of the other test stimuli (149 stimuli) for the single sound test samples. We performed pairwise identification for all the 149 pairs and defined the identification accuracy by the proportion of the instances where the presented stimulus showed a higher correlation coefficient. For the samples from the attention experiment, one candidate was the attended stimulus and the other was the unattended stimulus. For each attention sample, correct identification was defined as having a higher correlation with the attended stimulus.

\subsection*{Statistics}
All statistical tests were performed individually, treating each subject's results as within-subject replications of an experiment \cite{Ince2022Within-participant}. We planned to conduct a statistical analysis using a 95\% confidence interval to determine if the mean identification accuracy of the reconstructed sounds across test stimuli exceeded the chance level of 50\%. The sample size for the natural sound test experiment (\textit{N =} 50) was established prior to conducting the experiments, which are greater than the sample size required to detect the effect size of Cohen's \textit{d} = 0.5 at a significance level of 0.05 (\textit{N =} 27). For the attention experiment, we planned to conduct a binomial test to determine whether the proportion of correct identification exceeded the chance level of 50\%. The sample size  (\textit{N =} 48) was determined prior to the experiment, and is larger than the sample size required to detect the effect size of \textit{g} = 0.2 (correct rate = 0.7) at a significance level of 0.05 (\textit{N =} 37). Even though we used data samples from a 4-s time window for decoder training and reconstruction, statistical evaluations were performed on data points corresponding to 8-s stimulus blocks. This is to address the lack of independence among the three samples derived from an 8-s stimulus block. In the analyses of the single sound test samples, the identification accuracies of the three samples were averaged to define a single data point for statistical analysis, resulting in 50 data point for each condition and subject. In the analyses of the attention test samples, the binary outputs from the three samples of the same stimulus and attention condition were pooed by majority voting to define a single binary data point, resulting in 48 data points in each condition and subject. For the posthoc analyses using the ablated training category set, we could only use fewer data points for each condition than the predetermined size (\textit{N =} 10 for the animal, music, and environmental categories and \textit{N =} 20 for the speech category).

\subsection*{Author contributions}
JP: Conceptualization, Methodology, Software, Formal analysis, Investigation, Data curation, Writing - original draft, Writing - review \& editing, Visualization. M.Tsukamoto: Investigation, Writing - review \& editing. M.Tanaka: Investigation, Writing - review \& editing. YK: Conceptualization, Methodology, Validation, Resource, Writing - original draft, Writing - review \& editing, Supervision, Project administration, Funding acquisition.

\subsection*{Funding}
This research was supported by the KAKENHI grants from the Japan Society for the Promotion of Science (JSPS; https://www.jsps.go.jp), with grant numbers 20H05705 and 20H05954 assigned to YK. Additional financial support was provided by the New Energy and Industrial Technology Development Organization (NEDO; https://www.nedo.go.jp) under the grant number JPNP20006 to YK. Furthermore, YK also received backing through JST CREST (https://www.jst.go.jp/kisoken/crest/) with grant numbers JPMJCR18A5 and JPMJCR22P3. Despite this financial aid, the funders had no influence on the study's design, data collection, evaluation, the decision to publish, or the drafting of the manuscript.

\subsection*{Competing interests}
The authors declare no competing financial interests.

\subsection*{Acknowledgments}
This study was conducted using the fMRI scanner and related facilities of the Kyoto University Institute for the Future of Human Society. The authors would like to thank Eizaburo Doi, Fan Cheng, and Ken Shirakawa for their helpful discussion and comments on the manuscript.

\subsection*{References}
\bibliographystyle{unsrtnat}
\bibliography{refs.bib}


\onecolumn 
\fancyhead{} 
\renewcommand{\floatpagefraction}{0.1}
\lfoot[\bSupInf]{\dAuthor}
\rfoot[\dAuthor]{\cSupInf}
\newpage

\captionsetup*{format=largeformat} 
\setcounter{figure}{0} 
\setcounter{equation}{0} 
\makeatletter 
\renewcommand{\thefigure}{S\@arabic\c@figure} 
\makeatother
\def\theequation{S\arabic{equation}}


\newpage
\section*{Supplementary Information}

\label{S1_Video}
{\bf Supplementary Movie 1. Reconstructed sounds under natural sound listening conditions. } The movie presents a side-by-side comparison of the Mel-spectrograms of actual stimuli (left) and reconstructed sounds (right) for various natural sound stimuli. (Subjects: S3, S4, and S5, ROI: AC)
https://youtu.be/JgG\_HI8hXD4

\label{S2_Video}
{\bf Supplementary Movie 2. Reconstructed sounds using different auditory features.  } (Subject: S3, ROI: AC)
https://youtu.be/e4lzNyPFrmw

\label{S3_Video}
{\bf Supplementary Movie 2. Reconstructed sounds using brain decoder with ablated training category set.  } . (Subject: S3, ROI: AC)
https://youtu.be/lK9gN3LmUJU

\label{S4_Video}
{\bf Supplementary Movie 4. Reconstructed sounds using different reconstruction methods.  }  (Subject: S3, ROI: AC)
https://youtu.be/8qHH5Y9-X5g

\label{S5_Video}
{\bf Supplementary Movie 5. Reconstructed sounds using individual ROIs.   } (Subject: S3)
https://youtu.be/zMlzCWb0SWY

\label{S6_Video}
{\bf Supplementary Movie 6. Reconstructed sounds using different DNN layers.   } (Subject: S3, ROI: AC))
https://youtu.be/\_CJHJETO80k

\label{S7_Video}
{\bf Supplementary Movie 7. Reconstructed sounds under selective auditory attention tasks.  } On the left side, the Mel-spectrogram of the superimposed stimuli, derived from two distinct sound categories, is illustrated using different colors. On the right side, the Mel-spectrogram of the reconstructed sounds when the subject’s attention is focused on one of the two sounds (Subject: S4 and S5, ROI: AC)
https://youtu.be/8DkGngm82gA

\label{S8_Video}
{\bf Supplementary Movie 8. Reconstructed sounds under selective auditory attention tasks using individual ROIs.  } (Subject: S4 and S5)
https://youtu.be/33YA2FnyHNs

\newpage
    \includegraphics[width=.8\linewidth]{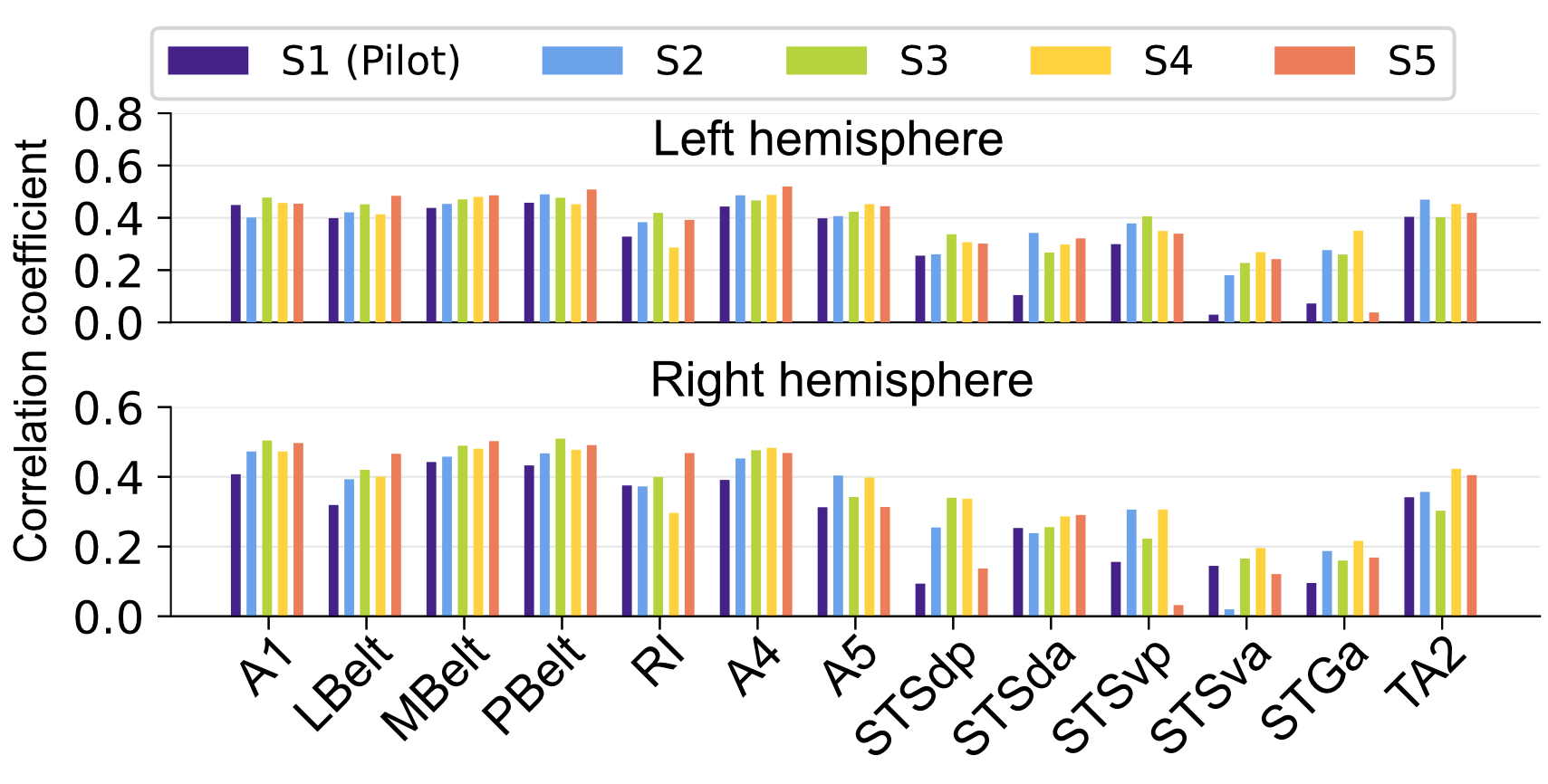}
\label{S1_Fig}
\newline
{\bf Supplementary Figure 1. Feature decoding performance of individual ROIs. } Decoding accuracy of DNN features of conv5 layer from VGGish-ish. Upper panel represents decoding accuracy of individual ROIs from left hemisphere, and lower panel represents decoding accuracy of individual ROIs from right hemisphere. Each bar represents the decoding accuracy of each subject by averaging across units.

\newpage
    \includegraphics[width=.8\linewidth]{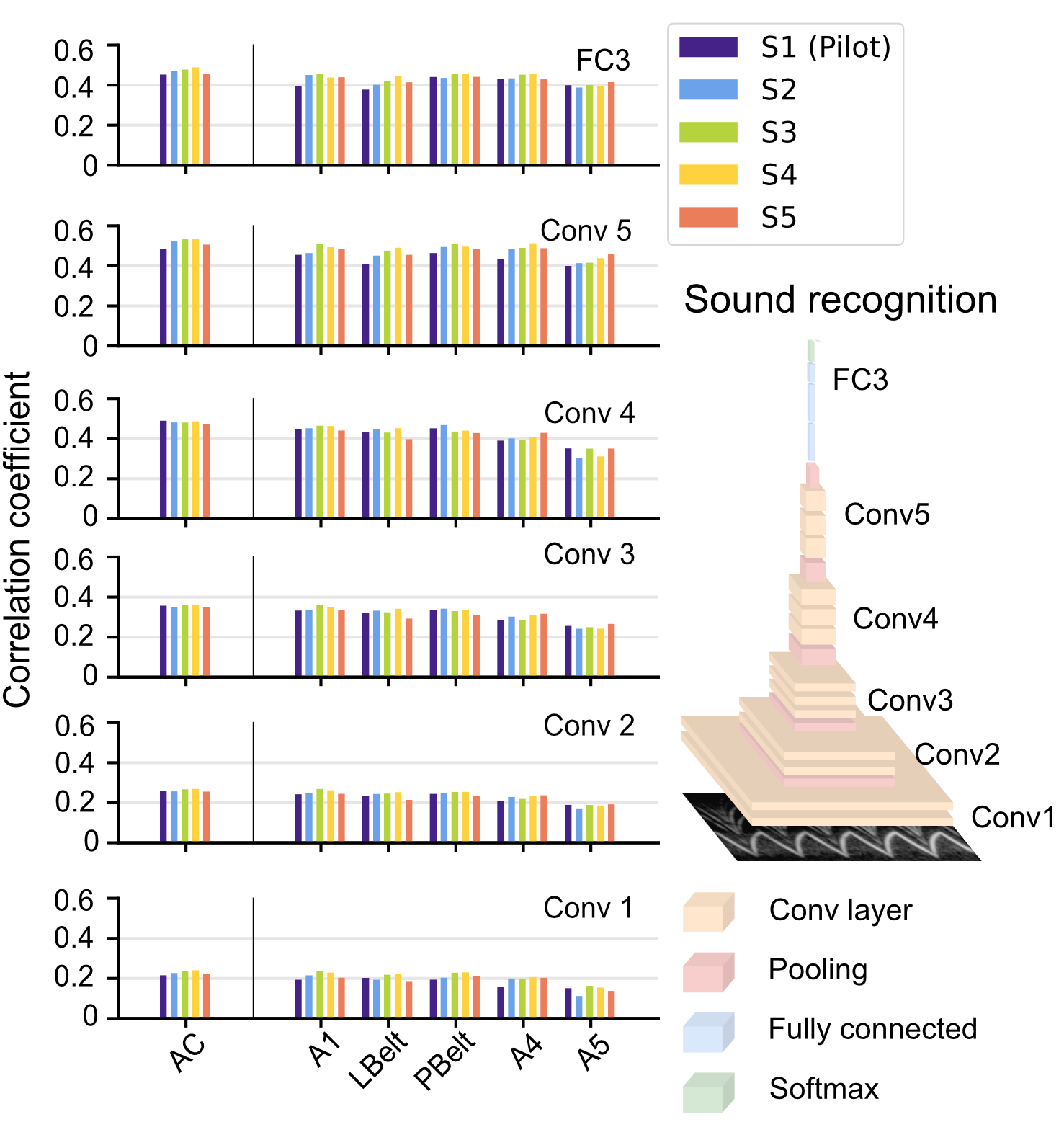}
\label{S2_Fig}
\newline
{\bf Supplementary Figure 2. Feature decoding performance of VGGish-ish layer. } Decoding accuracy of DNN features of six representative layers from sound classifier model. Each bar corresponds to the decoding accuracy of an individual subject, calculated as the average across all units for each layer.

\newpage
    \includegraphics[width=.8\linewidth]{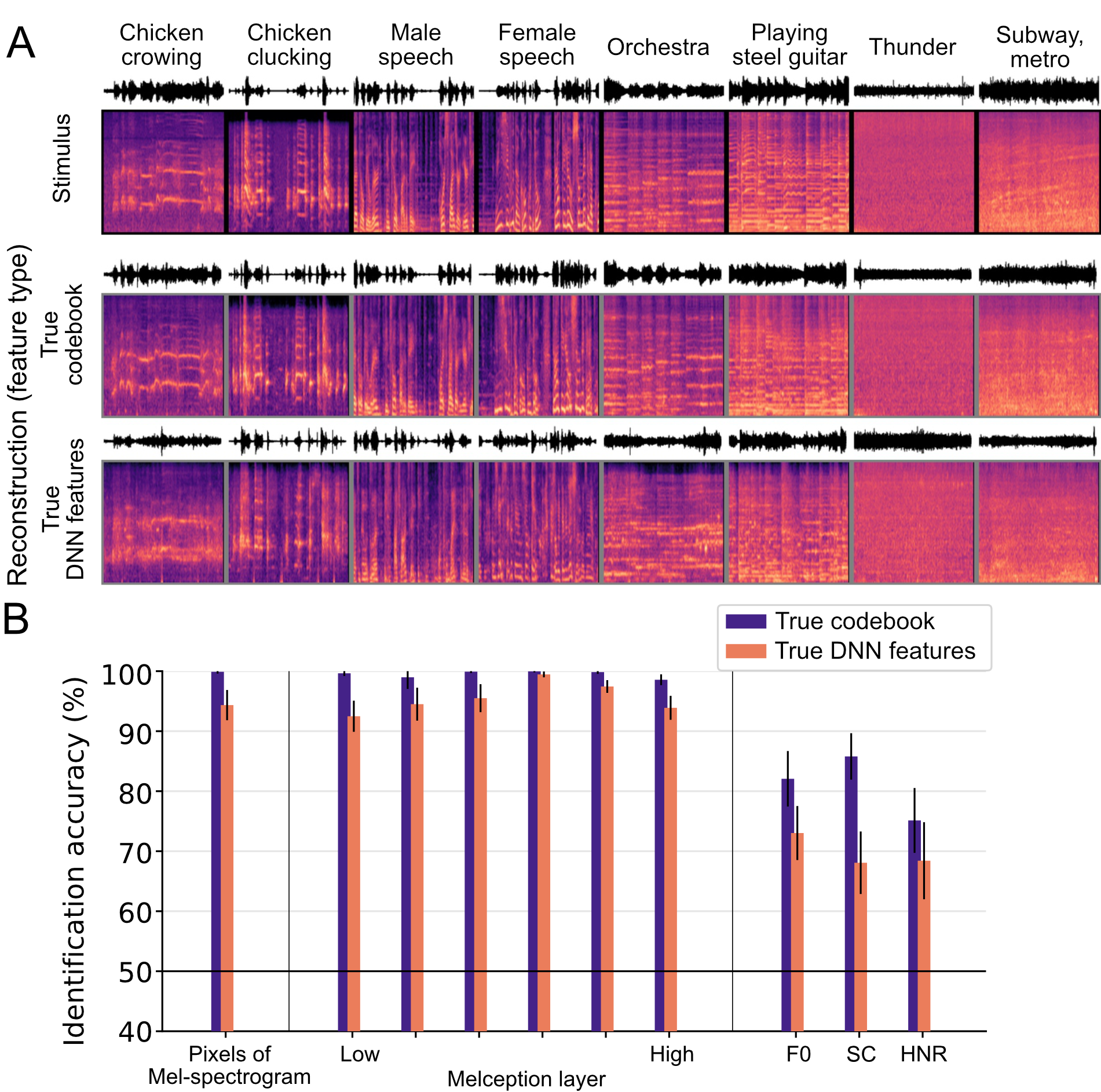}
\label{S3_Fig}
\newline
{\bf Supplementary Figure 3. Comparison of reconstructed sounds using true features. } (A) Reconstructed Mel-spectrogram using true codebook and DNN features. The first row displays the Mel-spectrogram of the actual sound, the second row shows the reconstructed Mel-spectrograms from true codebook representations, and the third row shows the reconstructed Mel-spectrogram from true DNN features. (B) Evaluation of the reconstructed sounds from true features. Each feature is represented by a different color.

\newpage
    \includegraphics[width=.8\linewidth]{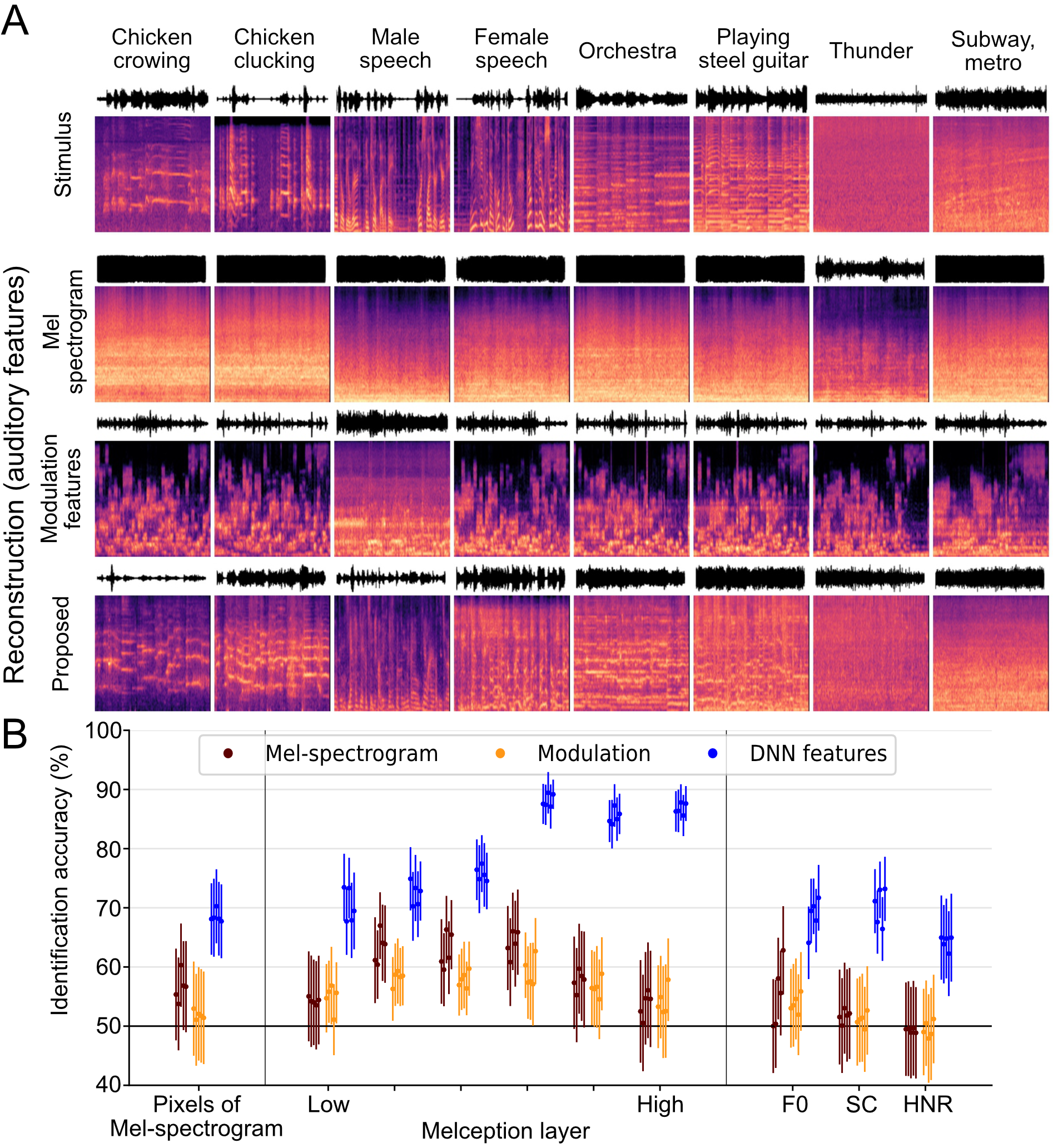}
\label{S4_Fig}
\newline
{\bf Supplementary Figure 4. Comparison of reconstructed sounds using different auditory features. }(A) Reconstructed Mel-spectrograms. The original Mel-spectrogram of the presented sound is displayed in the first row. The subsequent rows two to four show the reconstructed Mel-spectrograms using different auditory features from S3 using AC. (B) Evaluation of reconstruction sound using different auditory features. Each method is represented by a unique color. Each dot represents the mean identification accuracy calculated from each subject, with the error bar indicating the 95\% CI estimated using 50 data points.

\newpage
    \includegraphics[width=.8\linewidth]{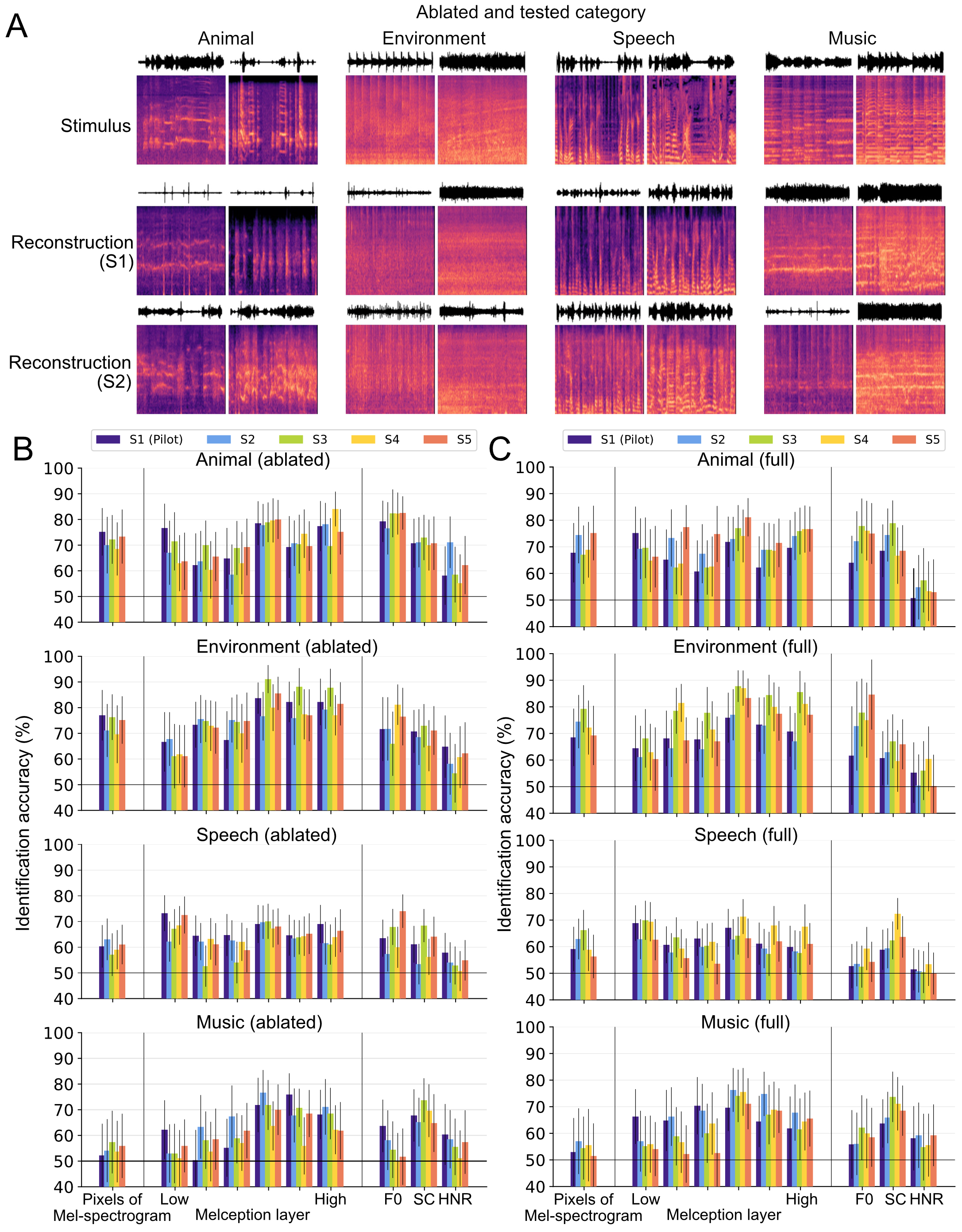}
\label{S5_Fig}
\newline
{\bf Supplementary Figure 5. Reconstruction with ablated training category sets. }  (A) Examples of reconstructed Mel-spectrograms of four categories (Environment, Animal, Speech, and Music) obtained with the decoders trained on the dataset where the data belonging to the same category was ablated. The first row displays the Mel-spectrogram of the presented sounds, and the second and the third rows show the reconstructed Mel-spectrograms from two subjects using AC and conv5. (B) Identification accuracies with the ablated training set. As in other analyses, the Mel-spectrogram pixels, hierarchical features of the Melception classifier, and the acoustic features were used for evaluation. Each bar represents the mean identification accuracy averaged across 10 test stimuli for Environment, Animal, and Music, and 20 test stimuli for Speech. The error bars indicate the 95\% confidence interval. Each subject is represented by a different color. (C) Identification accuracies with the full training set. The results of the same test data belonging to each category are displayed for comparison, in the same manner as in B.

\newpage
    \includegraphics[width=.8\linewidth]{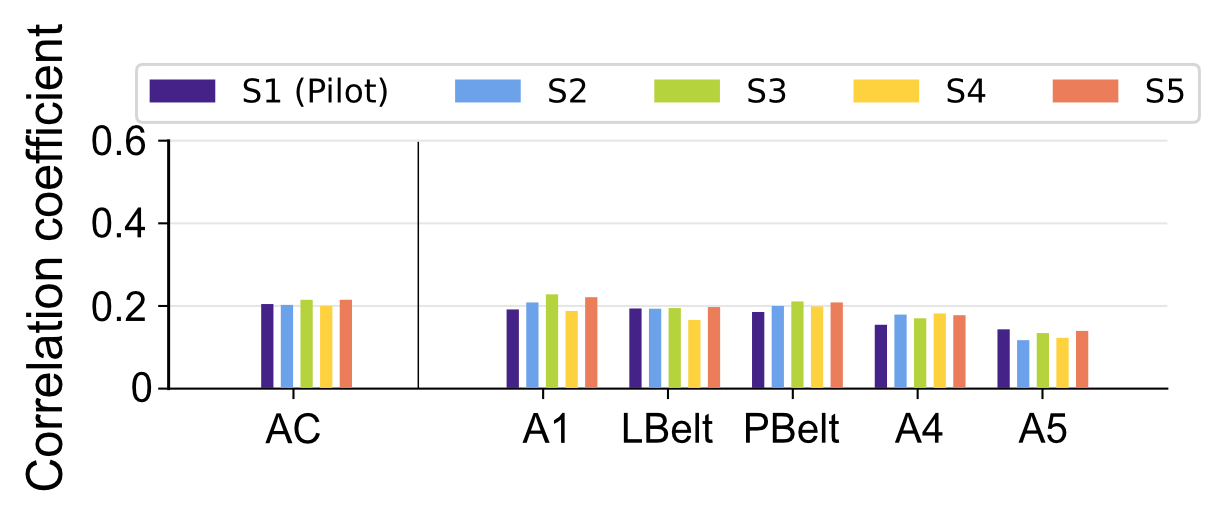}
\label{S6_Fig}
\newline
{\bf Supplementary Figure 6. Feature decoding performance of codebook representations. } Decoding accuracy of codebook representations. Each bar represents the decoding accuracy of each subject by averaging across units.

\newpage
    \includegraphics[width=.8\linewidth]{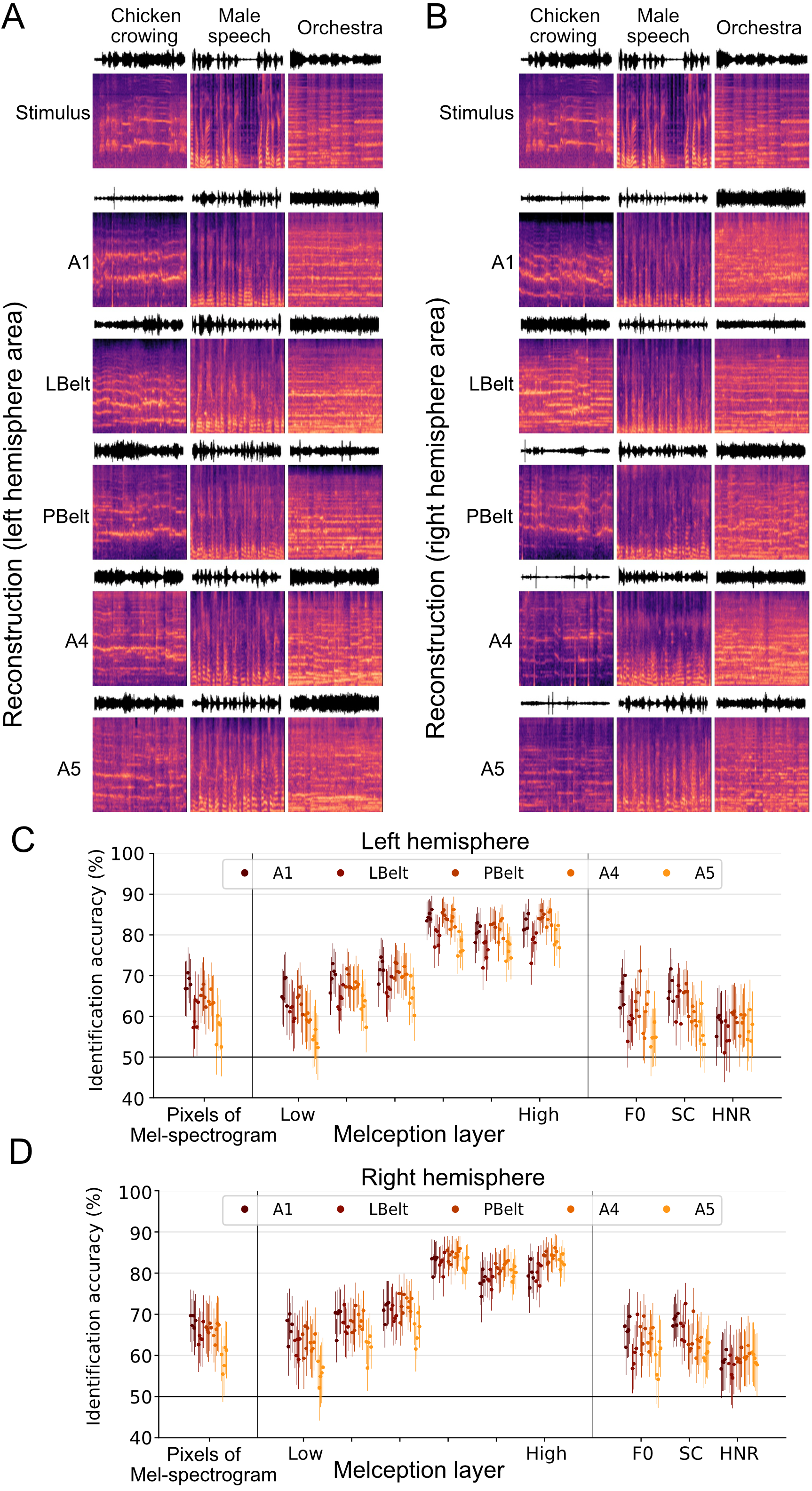}
\label{S7_Fig}
\newline
{\bf Supplementary Figure 7. Reconstructed sounds of all individual ROIs. } . (A) Reconstructed Mel-spectrogram using individual ROIs. Left panel represents reconstructed spectrogram from left hemisphere, and right panel represents reconstructed spectrogram from right hemisphere. The first row displays the Mel-spectrogram of the presented sound, and the second to the sixth row shows the reconstructed Mel-spectrograms from individual ROIs from S3 using conv5. (B) Evaluation of the reconstructed sounds from individual ROIs from left hemisphere. Each individual ROI is represented by a different color, and each dot shows the average identification accuracy across 150 test stimuli for each subject. (C) Evaluation of the reconstructed sounds from individual ROIs from right hemisphere, in the same manner as B.

\newpage
    \includegraphics[width=.75\linewidth]{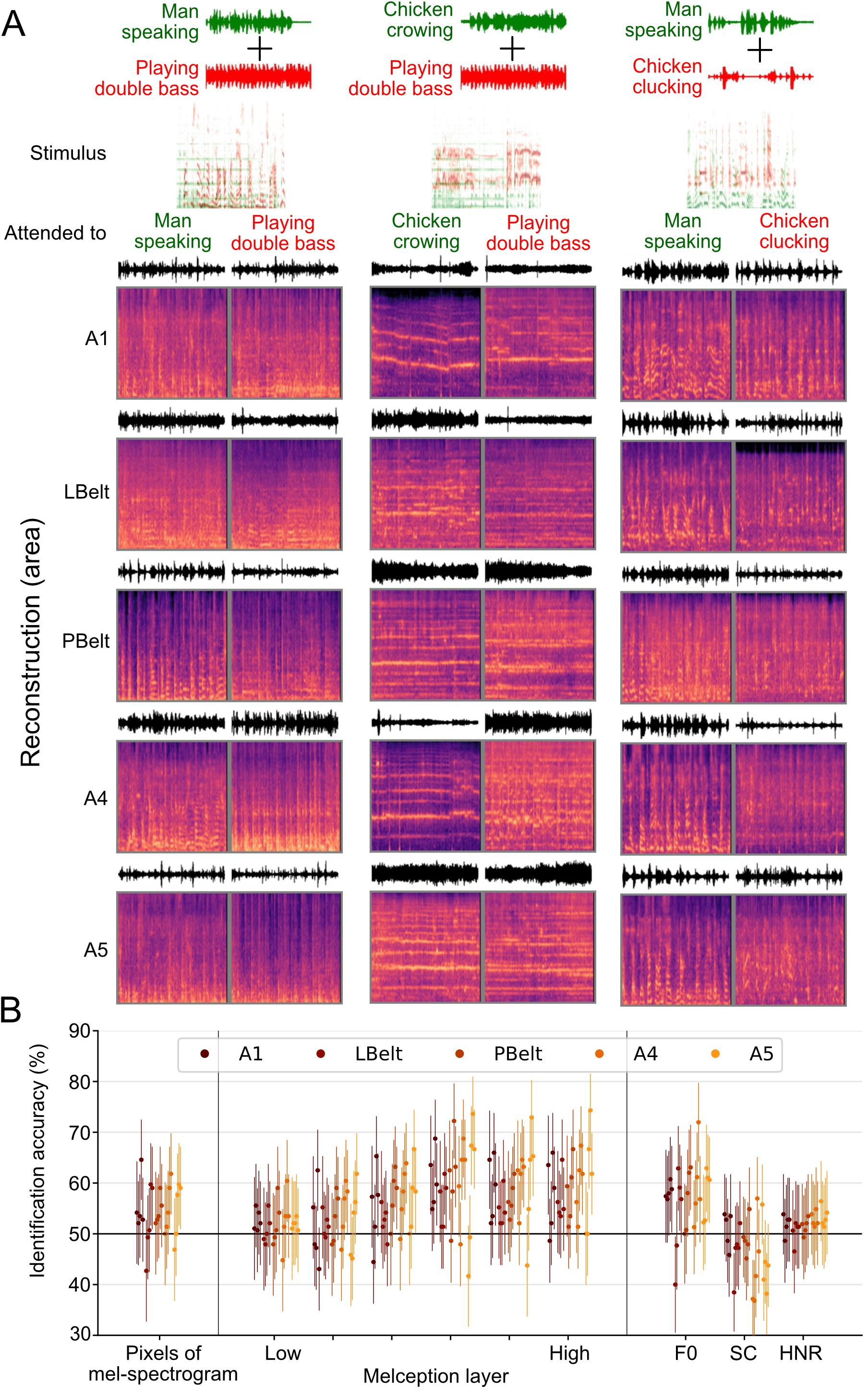}
\label{S9_Fig}
\newline
{\bf  Supplementary Figure 7. Reconstructed sounds of individual ROIs for selective auditory attention tasks. }(A) Reconstructed Mel-Spectrogram from Individual ROIs under selective auditory attention tasks. The top panel illustrates the Mel-spectrogram of the sound presented during the task. Here, subjects were instructed to focus on a specific sound from the superimposed sound stimuli. The bottom panel presents the Mel-spectrograms reconstructed from each individual ROI from S4 using conv5 (B) Evaluation of the reconstructed sounds from individual ROIs. Each individual ROI is represented by a different color. Each dot represents the mean identification accuracy calculated from each subject using 48 data points. The dashed line represents the accuracy level for statistical significance (\textit{p} < 0.05) as determined by the binomial test.

\end{document}